\documentclass[12pt]{iopart}

\usepackage{dcolumn}
\usepackage{bm}
\usepackage{epsf}
\usepackage{amsfonts}
\usepackage{amsmath}
\usepackage{amssymb}
\usepackage{pifont} 
\usepackage{float}
\usepackage[backref=page]{hyperref}
\usepackage{color}
\usepackage[dvipsnames]{xcolor}
\hypersetup{colorlinks, linkcolor={red!50!black}, citecolor={blue!50!black}, urlcolor={blue!80!black}
} 
\usepackage{graphicx}
\usepackage{subcaption}
\usepackage{dsfont}

\newcommand{\be}{\begin{equation}}
\newcommand{\ee}{\end{equation}} 
\newcommand{\eei}{\end{equation}\indent\indent}
\newcommand{\bc}{\begin{center}}
\newcommand{\ec}{\end{center}}
\newcommand{\ber}{\begin{eqnarray*}}
\newcommand{\ear}{\end{eqnarray*}}
\newcommand{\ba}{\begin{array}}
\newcommand{\ea}{\end{array}}

\newcommand{\bea}{\begin{eqnarray}}
\newcommand{\eea}{\end{eqnarray}}

\newcommand{\ei}{\end{itemize}}

\newcommand{\bra}[1]{\left(#1\right)}

\newcommand{\la}{\langle}
\newcommand{\ra}{\rangle}

\newcommand{\A}{{\cal A}}
\newcommand{\E}{{\cal E}}

\def\case#1/#2{\textstyle\frac{#1}{#2} }

\usepackage[normalem]{ulem}

\begin{document}

\title[Some exact relativistic star solutions in f(R) gravity]{Some exact relativistic star solutions in f(R) gravity}

\author{Mariam Campbell\textsuperscript{1}, Sante Carloni\textsuperscript{2,}\textsuperscript{3,}\textsuperscript{4}, Peter K. S. Dunsby\textsuperscript{1,}\textsuperscript{5,}\textsuperscript{6}, Nolene F. Naidu\textsuperscript{1}}
\address{$^1$Department of Mathematics and Applied Mathematics, Cosmology and Gravity Group,  University of Cape Town, Rondebosch, 7701, Cape Town, South Africa}
\address{$^2$DIME Sez. Metodi e Modelli Matematici, Universit\`{a} di Genova, Via  All’Opera Pia 15, 16145 - Genoa, (Italy).}
\address{$^3$INFN Sezione di Genova, Via Dodecaneso 33, 16146 Genova, Italy}
\address{$^4$Institute of Theoretical Physics, Faculty of Mathematics and Physics,
Charles University, Prague, V Hole{\v s}ovi{\v c}k{\' a}ch 2, 180 00 Prague 8, Czech Republic}
\address{$^5$South African Astronomical Observatory, Observatory 7925, Cape Town, South Africa}
\address{$^6$Centre for Space Research, North-West University, Potchefstroom 2520, South Africa}

\ead{CMPMAR009@myuct.ac.za}
\vspace{10pt}
\begin{indented}
\item[]\date{\today} 
\end{indented}

\begin{abstract}
We present a covariant description of non-vacuum static spherically symmetric spacetimes in $f(R)$ gravity applying the (1+1+2) covariant formalism. The propagation equations are then used to derive a covariant and dimensionless form of the Tolman-Oppenheimer-Volkoff (TOV) equations. We then give a solution strategy to these equations and obtain some new exact solutions for the particular case $f(R)=R+\alpha R^{2}$, which have the correct thermodynamic properties for standard matter. 
\end{abstract}
\noindent{\it Keywords }: Modified Gravity, TOV equations, exact solutions
%
%
%
%
%

\section{Introduction}
The remarkable discovery of the late-time accelerated expansion of the Universe has, over the past two decades, led to numerous studies of different models of the Dark Universe aimed at overcoming the limitations of the standard $\Lambda$CDM cosmological model. 

One of the most popular alternatives to the standard model is based on gravitational actions, which are nonlinear in the Ricci scalar $R$  --- the so-called $f(R)$ theories of gravity. This is because the non-linear corrections to the Hilbert-Einstein action can be recast as effective fluid quantities, which naturally lead to violations of the strong energy condition and consequently accelerated expansion without introducing additional fields. Such models first became popular in the 1980s because it was shown that they can be derived from fundamental physical theories (for example, M-theory) and naturally admit a phase of accelerated expansion associated with inflation in the Early Universe \cite{Buchdahlmodgrav,Starobinsky:1980te}. The fact that Dark Energy requires the presence of a similar phase of accelerated expansion at late times has revived interest in these theories and led to a considerable amount of work, both in cosmological and astrophysical applications \cite{fRPerturbations}.

Because the number of potential $f(R)$ candidate theories is large, there needs to be a systematic comparison between all theoretical predictions of a given theory with the available cosmological data sets (Cosmic Microwave Background, Large Scale Structure, Baryon Acoustic Oscillations, Type Ia Supernova, etc.). A somewhat better approach is not to assume the form of the gravitational action but rather attempt to constrain it from cosmological data, assuming that the  Copernican principle holds. Such a cosmographic approach has the advantage of being model-independent. Unfortunately, all of these procedures suffer to some degree from the so-called degeneracy problem, i.e., several competitive gravitational theories are consistent with the available data at the same statistical precision. 

One way to address this problem and further constrain the number of experimentally viable $f(R)$ theories is to improve our understanding of their phenomenology and limitations in other contexts, like, e.g., astrophysics. This provides a way of probing the high-energy (or strong gravitational) limit of these theories, which is complementary and perhaps even more strongly motivated than the picture obtained from the low-curvature, late-time cosmological evolution. Of particular interest are investigations of the existence and properties of relativistic compact objects, such as white dwarfs and neutron stars, and gravitational collapse (see \cite{adrian-collapse} as an example of its application in $f(R)$ gravity).  

In particular, the development of a description of relativistic stars involves a detailed study of the Tolman-Oppenheimer-Volkoff (TOV) equation. Introduced in 1939 \cite{TOV}, these equations provide a way of determining the pressure profile of a static, spherically symmetric object in General Relativity (GR). Although obtaining exact solutions of the TOV equations is a formidable task, some solutions do exist; see, e.g., in \cite{GR-solutions}. 

The situation becomes considerably more complicated in $f(R)$ gravity because the field equations are fourth order. Until now, all studies of relativistic stars in these theories involve numerical integration of the governing equations. As far as we know, no non-vacuum exact solutions for stars exist \cite{numerical}. An extensive review of compact stellar objects in extended theories of gravity is covered in \cite{Olmo-Garcia-Wojnar-review}.

Recently, a new approach to the treatment of the TOV equations has been proposed using covariant semi-tetrad methods called the (1+1+2) approach. Developed by Clarkson and Barrett \cite{extension}, this approach has been recently applied to study Schwarzschild black holes, such as their linear perturbations \cite{extension} and the production of a stream of electromagnetic radiation that mirrors the black hole ring-down when gravitational waves around a vibrating Schwarzschild black hole interact with a magnetic field \cite{Betschart}. 

Applications of this formalism to the problem of modeling the interior of isotropic relativistic stars in GR were proposed in
\cite{sante-TOV-iso-GR1,Luz:2019frs} and extended to a two-fluid framework by \cite{NCD}.  However, more realistic modeling requires the introduction of anisotropies. Indeed, anisotropies arise naturally in astrophysical systems, for e.g., gravitational collapse, and rotating stars such as pulsars \cite{anisogravitationalcollapse,pulsarsdynamicsandstructure,anisonuetronstars} including high-density compact objects \cite{high-density-compact-objects}. Compact anisotropic stars that describe realistic astrophysical phenomena, such as neutron stars, have been explored in GR \cite{MakHarko}. The consequence of local anisotropy in self-gravitating systems in Newtonian and general relativistic cases causes a non-negligible effect on the critical mass of a stellar object, whereby it is more or less stable compared to the local isotropic case \cite{self-gravitating-aniso}. Anisotropic compact objects have also been analyzed in covariant frameworks by \cite{sante-TOV-aniso-GR1,anisostarsGR,Luz:2024yjm,Luz:2024xnd,Luz:2024lgi}, and extended to a two-fluid system in GR \cite{NCD2}, making it well-suited to extended models of GR. In particular, it was shown how to generate two fluid solutions either through direct resolution or by reconstructing them from known single fluid solutions \cite{NCD,NCD2}.

Our study focuses on compact objects within $f(R)$ theories of gravity. It is well known that anisotropies play an important role in these theories. As an example, De Felice and Tsujikawa, and Sotiriou and Faraoni \cite{Buchdahlmodgrav}, discuss how anisotropy arises in charged and/or rotating black holes in $f(R)$ gravity, and Nashed and Capozziello show how anisotropic compact stars in $f(R)$ gravity can describe realistic systems such as pulsars \cite{fR-compact-object}.  Since $f(R)$ gravity naturally introduces anisotropy we will assume that also the fluid sources are anisotropic. This choice is motivated by generality but also by the fact it is comparatively easier to obtain a solution sourced by a more general form of matter.

To exploit the symmetry of our problem, the (1+1+2) covariant formalism is employed, and in the context of $f(R)$ gravity, this approach has been used to describe a spherically symmetric vacuum solution in $f(R)$ gravity \cite{Nzioki}. The same authors also studied the gravitational lensing properties of spherically symmetric spacetimes in $f(R)$ gravity \cite{Nzioki:2010nj}. Another example of the advantages of working in this formalism is the study \cite{sante-scalar-tensor} where the authors could easily show, in a coordinate independent way, that no scalar-tensor theory of gravity admits a Schwarzschild solution unless one considers a trivial scalar field.

In this paper, we formulate and solve exactly the TOV equations for $f(R)$ gravity. More specifically, we use the fact that $f(R)$ theories can be written as GR plus baryonic matter and an effective {\it``curvature fluid"}. This allows us to use the methods developed in \cite{NCD,NCD2} to generate exact solutions of the TOV equation in $f(R)$ gravity. Adopting the (1+1+2) covariant formalism \cite{extension} and the methods used in \cite{NCD,NCD2}, for the first time, we are able to write down analytical solutions to the TOV equations in the context of $f(R)$ gravity.

The outline of this paper is as follows. In Sec. \ref{sec:FE}, we present the field equations of $f(R)$ gravity. In Sec. \ref{sec:covariant}, we review the (1+1+2) covariant semi-tetrad formalism and specialize the equations to locally rotationally symmetric (LRS) type II spacetimes. We then apply this formalism to $f(R)$ gravity in Sec. \ref{sec:lrs}. In Sec. \ref{sec:tov}, we derive the TOV equations for a two-fluid system, which is well suited to the study of spherically symmetric non-vacuum solutions in $f(R)$ gravity. In Sec. \ref{sec:physbound}, we turn to the important issues of thermodynamical constraints for matter sources and junction conditions. These conditions will be needed to properly match our solutions to the exterior vacuum Schwarzschild geometry. At this point, we are ready to use our formalism to generate new solutions. 
In Sec. \ref{sec:S-T}, we define the reconstruction algorithm needed to obtain the exact solutions, and in Sec. \ref{sec: CD-TIV}, we construct and examine some new solutions for a quadratic $f(R)$ model. Finally, we present our conclusions in Sec. \ref{sec:conclusion} and discuss possible future work.

To close off this section, we provide a few standard definitions and conventions that will be used throughout this paper. Natural units will be used ($\hbar=c=k_{B}=8\pi G=1$). The covariant derivative and partial differentiation are denoted by the symbols $\nabla$ and $\partial$, respectively, and Latin indices are used for space (1-3 indices) and time (0 index) components. The metric signature $-,+,+,+$ is used. 
The Riemann tensor is defined by
\begin{equation}
R^{a}{}_{bcd}=\Gamma^a{}_{bd,c}-\Gamma^a{}_{bc,d}+ \Gamma^e{}_{bd}\Gamma^a{}_{ce}-\Gamma^{e}_{bc}\Gamma^a{}_{de}\;,
\end{equation}
where the metric connection $\Gamma^a{}_{bd}$ is the Christoffel symbols, given by
\begin{equation}
\Gamma^a_{bd}=\frac{1}{2}g^{ae}
\left(g_{be,d}+g_{ed,b}-g_{bd,e}\right)\;.
\end{equation}
The Ricci tensor is defined as the contraction of the {\em first} and the {\em third} indices of the Riemann tensor
\begin{equation}\label{Ricci}
R_{ab}=g^{cd}R_{acbd}\;.
\end{equation}
A tensor that is symmetric and antisymmetric on the indices is defined as 
\begin{equation}
T_{(a b)}= \frac{1}{2}\left(T_{a b}+T_{b a}\right)\;,\qquad T_{[a b]}= \frac{1}{2}\left(T_{a b}-T_{b a}\right)\,,
\end{equation}
respectively.
Finally, in standard GR, including a matter field, the Einstein-Hilbert action is 
\begin{equation}
{\mathds A}=\frac12\int d^4x \sqrt{-g}\left[R+ 2{\cal L}_m \right]\;.\label{GR action}
\end{equation}  

\section{The Field Equations} \label{sec:FE}

A general description of a fourth-order theory of gravity includes the introduction of additional curvature invariants, such as $R$, $R_{ab}R^{ab}$ and $R_{abcd}R^{abcd}$, to \eqref{GR action}. One of the simplest possible generalizations of this kind of theories, which turns out to be fairly general in four dimensional spacetimes with high symmetry, \cite{DeWitt:1965jb,Barth:1983hb}, is given by the action
\be
{\mathds A}= \frac12 \int d^4x\sqrt{-g}\left[f(R)+2{\cal L}_m\right]\;,
\label{action}
\ee
where ${\cal L}_m$ describes the matter field.

The general modified field equations are obtained by varying \eqref{action} with respect to the metric $g_{ab}$:
\be
f' R_{ab}- \frac12 f g_{ab}
- \nabla_{b}\nabla_{a}f' + g_{ab}\nabla_{c}\nabla^{c}f' = T^{m}_{ab}, 
\label{field0}	
\ee
where $T^{m}_{ab}$ represents the stress-energy tensor of the matter sources, $f\equiv f(R)$, and $f'\equiv df/dR$. The above equation can be recast as
\be
G_{ab} =T^{\textbf{eff}}_{ab} = \tilde{T}^{m}_{ab}+ T^{R}_{ab}, 
\label{field1}	
\ee
where 
\be
T^{R}_{ab}\equiv\frac{1}{f^{\prime}}\left[\frac12 (f-Rf') g_{ab}
+ \nabla_{b}\nabla_{a}f'- g_{ab}\nabla_{c}\nabla^{c}f'\right],
\label{emt R}
\ee
and 
\be
\tilde{T}^{m}_{ab}\equiv T^{m}_{ab}/f^{\prime}.
\ee
Expressing the field equations, Eq. \eqref{field0}, in the form of Eq. \eqref{field1} allows us to consider higher order corrections to the Einstein field equations as an effective fluid, thus providing a way to employ some of the results in GR  (and of \cite{NCD,NCD2}) to find analytical two-fluid interior solutions to compact objects.

In $f(R)$ gravity, the trace of the field equations, Eq. \eqref{field1},
\be
\begin{split}
 R=&\frac{1}{f^{\prime}}(3p^{m}-\mu^{m}) + \frac{2f}{f^{\prime}} - 3\frac{f^{\prime\prime\prime}}{f^{\prime}}\nabla^{a}R\nabla_{a}R- 3\frac{f^{\prime\prime}}{f^{\prime}}\nabla^{2}R
        + 3\Theta\dot{R}\frac{f^{\prime\prime}}{f^{\prime}}+\\
        &+ 3\frac{f^{\prime\prime}}{f^{\prime}}\ddot{R} + 3\frac{f^{\prime\prime\prime}}{f^{\prime}}\dot{R}^{2} - 3\dot{u}^{c}\frac{\left(\nabla_{c}f^{\prime}\right)}{f^{\prime}},\label{eqn: trace}
\end{split}
\ee
will prove to be particularly important in writing down the modified TOV equations. It captures the dynamics of the additional scalar degree of freedom that characterizes $f(R)$ theories.

The twice contracted Bianchi identities tell us that the divergence of the left-hand-side of Eq. \eqref{field1} is identically zero. Hence, the right-hand-side will be zero resulting in $T^{\textbf{eff}}_{ab}$ being conserved. This leads to an important consequence: if baryonic matter is conserved, the total fluid is also conserved. However, it should be noted that this consequence does not imply that the individual fluids are conserved, i.e.,
\be
    \nabla^{b}\left(\frac{T^{m}_{ab}}{f^\prime}\right)=-\nabla^{b}T^{R}_{ab}=-\frac{f^{\prime\prime}}{f^{\prime2}}T^{m}_{ab}\nabla^{b}R.
\ee

We would also like to emphasize that $T^{R}_{ab}$ and $\tilde{T}^{m}_{ab}$ in Eq. \eqref{field1} both represent an \textit{effective} fluid. This means it could present unphysical properties for a fluid composed of baryonic matter. In analyzing the solutions presented in the proceeding sections, we will make sure that $T_{ab}^{m}$ satisfy several conditions that guarantee that the source fluid is physical, but allow for  $T^{R}_{ab}$ and $\tilde{T}^{m}_{ab}$ to have unphysical values.

Among all the possible forms of the function $f$, a particularly interesting choice is a quadratic polynomial.  In this case, we have a gravitational action in which a quadratic Ricci scalar term is added to the Einstein-Hilbert action:
\begin{equation}
{\mathds A}=\frac12\int d^4x \sqrt{-g}\left[R+ \alpha R^2+ 2{\cal L}_m \right]\;.\label{Star action}
\end{equation}
If the constant $\alpha$ is positive,  this model is called the {\it Starobinsky model}. Initially, this model was proposed as an effective action, representing quantum corrections in the matter content of spacetime. In a cosmological setting, Starobinsky showed that his model could induce an inflationary phase without the need to introduce a scalar field \cite{Starobinsky:1980te}. This theory is also proven to be ghost-free when deriving the particle spectrum of the theory, a feature that is rare in $f(R)$ gravity (see \cite{Buchbinder:2021wzv} for an introduction to this specific issue).  For our purposes, an important property of this model is that the only static spherically symmetric asymptotically flat solution with a regular horizon for this model is the Schwarzschild solution \cite{starobinskyuniquenesstheorem}. Consequently, such a model naturally contains an ideal representation of the exterior of a compact object, and, also for this reason,  it has been extensively studied in spherically symmetric spacetimes in modified theories of gravity \cite{starobinsky-stars}. In the next sections, we will use a quadratic model of gravity, Eq. \eqref{Star action}, where $\alpha$ is a free parameter. This will allow us to explicitly explore the corrections induced by unique structures that arise in these models of gravity, called double layers.

Astrophysical tests of gravitational interactions place constraints on $df/dR$ for a general $f(R)$ theory. Currently, the galactic halo sets the strongest bound $|f'|\leq10^{-6}$ \cite{lssconstraintfR}. Solar system tests, like the geodetic precession of an orbiting gyroscope around Earth, place an upper bound on the scalar curvature $R\leq10^{-22}m^{-2}$ \cite{Rconstraint}, and Mercury's precession rate bounds the parameter $\alpha$ as $|\alpha|\leq10^{18}~m^{2}$ \cite{mercuryconstraint}. The bounds on the parameter $\alpha$ remain inconclusive since the Gravity Probe B experiment and the binary pulsar system PSR J0737-3039 set a constraint on the parameter $\alpha$ as $5\times10^{11}$ $m^{2}$ \cite{gravityprobeB} and $2.3\times10^{15}$ $m^{2}$ \cite{binarypulsarconstraint, Rconstraint} respectively. Still, the E\"{o}t-Wash laboratory experiment gives an upper bound on $\alpha$ as $\alpha\leq10^{-10}$ $m^{2}$ \cite{eot-wash-lab-test}. Therefore, when considering quadratic models of $f(R)$, these constraints do not limit the parameter $\alpha$ since $1+2|\alpha|R\leq10^{-6}$. In addition to these limits, Ref. \cite{ConstraintsonfRviaNeutronStars} place a bound on the quadratic model parameter $|\alpha|\sim10^{9}\;cm^{2}$ by considering realistic equations of state for neutron stars. 
Frameworks for discriminating between extended models of gravity using gravitational waves have been investigated by \cite{ObservationsfR}. However, the parameter constraint on quadratic gravity remains contentious since studies on gravitational wave emissions from inspiralling black holes find $\alpha\sim10^{31}~m^{2}$ \cite{GWBHconstraint1} and $\alpha\leq1.1\times10^{13}~m^{2}$ \cite{GWBHconstraint2}.

\section{The (1+1+2) covariant formalism} \label{sec:covariant}
The (1+3) covariant approach, developed by Ehlers and Ellis \cite{Covariant}, has been instrumental in cosmological applications such as studying perturbation theory \cite{Perturbations} and CMB anisotropies \cite{CovCMB}. This approach is well suited to investigate cosmological spacetimes. For example, it can describe fully anisotropic but spatially homogeneous spacetimes (Bianchi models) via a set of ordinary differential equations comprised of scalar variables. The (1+3)  approach relies on a threading of the spacetime with the introduction of a time-like vector field $u^{a}$. This vector allows to define of a set of three-dimensional hypersurfaces (orthogonal to $u^{a}$) whose geometry is  described  by
\be
h_{ab}=g_{ab}+u_{a}u_{b}.
\ee
One can also define a derivative operator along $u^{a}$, which is given, for a generic tensor $\psi_{a...b}$, by
\be
\dot{\psi}_{a...b}\equiv u^{d}\nabla_{d}\psi_{a...b},
\ee
and a derivative on the 3-surfaces
\be
{\rm D}_{c}\psi_{a...b}\equiv h_{c}{}^{d}h_{a}{}^{e}...h_{b}{}^{f}\nabla_{d}\psi_{e...f}.
\ee
All the physical and geometrical descriptions are captured in kinematic and dynamic variables, which satisfy evolution and constraint equations derived from the Bianchi and Ricci identities \cite{Covariant}.

Our study will employ an extension of the (1+3) formalism, called (1+1+2) covariant approach \cite{extension},  which is obtained by a further threading of the 3-space defined by $h_{ab}$. In particular, a unit vector $e^{a}$ that is orthogonal to the 4-velocity $u^{a}$ is introduced, such that
\be
e_{a} u^{a} = 0\;,\; \quad e_{a} e^{a} = 1.
\ee
Then, the 2-surfaces geometry is characterized by
\be 
N_{ab} \equiv h_{ab} - e_{a}e_{b} = g_{ab} + u_{a}u_{b} 
- e_{a}e_{b}~,~~N^{a}{}_{a} = 2~, 
\label{projT} 
\ee 
which is orthogonal to $e^{a}$ and $u^{a}$. 

For the study of non rotating relativistic stars, it is sufficient to focus on the use of this approach in \textit{locally rotationally symmetric} (LRS) spacetimes and, more specifically, to the static LRS-II subclass, which is rotation-free. This class of spacetimes has the remarkable property that all the (1+1+2) quantities necessary for their description are scalars. 

In particular, given a  3-vector $v^{a}$ and a \textit{projected symmetric trace free} (PSTF) 3-tensor $\psi_{ab} $, we have
\be
v^{a} =  V e^{a}\,, \quad V\equiv v^{a} 
e_{a}\,, 
\label{equation1} 
\ee 
\be 
\psi_{ab} = \psi_{\la ab\ra} = \Psi\bra{e_{a}e_{b} - \frac{1}{2}N_{ab}}. 
\label{equation2} 
\ee 

In order to fully describe the propagation of the (1+1+2) quantities, we need to define, other than the derivative along  $u^{a}$, the derivatives along $e^{a}$ and on the 2-surface:
\bea
\hat{\psi}_{a..b}{}^{c..d} &\equiv & e^{f}D_{f}\psi_{a..b}{}^{c..d}~, 
\\
\delta_f\psi_{a..b}{}^{c..d} &\equiv & N_{a}{}^{f}...N_{b}{}^gN_{h}{}^{c}..
N_{i}{}^{d}N_f{}^jD_j\psi_{f..g}{}^{i..j}\;.
\eea 
In static LRSII spacetimes, the key quantities needed to describe the geometry are
\bea 
\mathcal{A}&\equiv &e^{a}\dot{u}_{a},\\
\phi &\equiv & \delta_ae^a~,\\
\E &\equiv & C_{acbd}u^{c}u^{d}e^{a}e^{b},
\eea
where $\mathcal{A}$ represent the acceleration of the observers that move with velocity $u^{a}$, $\phi$ describes the 2-surfaces expansion and $\E$ the electric part of  is the Weyl tensor $C_{acbd}$.

In addition to the (1+1+2) variables above, the complete set includes the variables resulting from the thermodynamics of the source fluid. These variables are obtained by the decomposition of the energy-momentum tensor of the matter fields, whose most general form, compatible with LRS-II spacetimes, is: 
\be \label{STEgen}
T_{ab}^{\textbf{tot}}=\mu^{\textbf{tot}}u_{a}u_{b} + (p^{\textbf{tot}}+\Pi^{\textbf{tot}})e_{a}e_{b} 
+ \left(p^{\textbf{tot}}-\frac{1}{2}\Pi^{\textbf{tot}}\right)N_{ab} + 2Q^{\textbf{tot}}e_{(a}u_{b)},
\ee
 where $\mu^{\textbf{tot}}$ is the total energy density of baryonic matter, $p^{\textbf{tot}}$ is the total isotropic pressure of baryonic matter, $q^{\textbf{tot}}_{a}$ is the total energy flux of baryonic matter, and $\pi^{\textbf{tot}}_{ab}$ is the total PSTF anistropic stress.  The apex ``$\textbf{tot}$'' in the above formula represents the fact that in the presence of more than one matter source, those quantities can be written as the sum of the individual fluids, i.e., in the case of two fluids $\mu^{{\textbf{tot}}}=\mu_{1}+\mu_{2}$, $p^{{\textbf{tot}}}=p_{1}+p_{2}$, and $\Pi^{{\textbf{tot}}}=\Pi_{1}+\Pi_{2}$. 

We now have all the fundamental quantities that describe our spacetime in the (1+1+2) formalism. Restricting our study to the case of static spherically symmetric LRS-II spacetimes, the two-fluid propagation equations are \cite{NCD2}
\bea
&&\hat\phi = -\frac12\phi^2 -\frac23\mu^{{\textbf{tot}}}-\frac12\Pi^{{\textbf{tot}}}-\E~,
\label{equation1a}\\
&&\hat\E -\frac13\hat\mu^{{\textbf{tot}}} + \frac12\hat\Pi^{{\textbf{tot}}} =- \frac32\phi\bra{\E+\frac12\Pi^{{\textbf{tot}}}}~,
\label{equation2a}\\
&&0 = - \A\phi + \frac13 \bra{\mu^{{\textbf{tot}}}+3p^{{\textbf{tot}}}} -\E +\frac12\Pi^{{\textbf{tot}}}~,
\label{equation3a}\\ 
&&\hat p^{{\textbf{tot}}}+\hat\Pi^{{\textbf{tot}}}= -\bra{\frac32\phi+\A}\Pi^{{\textbf{tot}}}-\bra{\mu^{{\textbf{tot}}}+p^{{\textbf{tot}}}}\A~,
\label{equation4a}\\
&&\hat\A = -\bra{A+\phi}\A + \frac12\bra{\mu^{{\textbf{tot}}} +3p^{{\textbf{tot}}}}~, 
\label{equation5a} 
\eea
together with the Gaussian curvature constraint
\be
K = \frac{1}{3}\mu^{{\textbf{tot}}}-\E-\frac{1}{2}\Pi^{{\textbf{tot}}}+\frac{1}{4}\phi^{2}.
\ee
We will now apply the formalism above to the case of $f(R)$ gravity.

\section{Static, spherically symmetric f(R) equations} \label{sec:lrs}
As previously mentioned, an advantageous feature of $f(R)$ theories of gravity is that one can express the field equations in such a way that it resembles GR with a two-fluid source comprised of non-minimally coupled matter and an effective \textit{curvature} fluid \cite{fRPerturbations}.
Therefore, our set-up is analogous to the two-fluid construction of the previous section, and therefore, the (1+1+2) equations in this case can be obtained by simply setting
\begin{equation}
T_{ab}^{\textbf{tot}}=T^{\text{eff}}_{ab}
\end{equation}
 in Eqs. (\ref{equation1a}-\ref{equation5a}), or, equivalently, by choosing
\bea
    &&\mu^{\textbf{tot}}=T^{\textbf{eff}}_{ab}u^{a}u^{b}=\frac{\mu^{m}}{f^{\prime}}+\mu^{R},\\
    &&p^{\textbf{tot}}=\frac13T^{\textbf{eff}}_{ab}\bra{e^{a}e^{b} +2N^{ab}}=\frac{p^{m}}{f^{\prime}}+p^{R},\\
    &&\Pi^{\textbf{tot}}=\frac23T^{\textbf{eff}}_{ab}\bra{e^{a}e^{b} -N^{ab}}=\frac{\Pi^{m}_{ab}}{f^\prime}+\Pi^{R}_{ab},\\
    &&Q^{\textbf{tot}}=-\frac{1}{2}T^{\textbf{eff}}_{bc}u^{c}e^{b}=-\frac{Q^{m}}{f^\prime}+Q^{R},
\eea
where the curvature quantities are defined as
\bea
\mu^{R} &=& \frac{1}{f'}\left(\frac12 ( Rf' - f) + f''\hat X + f''X \phi+ f'''X^{2}\right),\label{eq:muR}\\
p^{R} &=& \frac{1}{f'}\left(\frac12 ( f - Rf') - \frac{2}{3} f''\hat X - \frac{2}{3} f''X \phi -\frac{2}{3} f'''X^{2}-\A f''X\right),\label{eq:pR}\\
\Pi^{R}&=&\frac{1}{f^{\prime}}\left(\frac{2}{3}f^{\prime\prime}\hat{X} + \frac{2}{3}f^{\prime\prime\prime}X^{2} - \frac{1}{3}f^{\prime\prime}X\phi\right),
\label{eq:piR}\\
Q^{R}&=& -\frac{1}{f^{\prime}}\left(f'''\dot{R}X+f''(\dot{X}-\mathcal{A}\dot{R})\right)= 0, 
\eea
and $\hat{R}\equiv X$. Using the covariant formalism and the variables above, the trace equation, Eq. \eqref{eqn: trace}, can be written as $Rf=3p^{\textbf{eff}} - \mu^{\textbf{eff}}$ or
\be
Rf'-2f = 3p^{m} - \mu^{m}- 3f''\hat X - 3f''X \phi +- 3f'''X^{2} - 3\A f''X \;.\label{trace1}
\ee
For our purposes, a more useful form of the trace equation is
\be
\hat X = \frac{p^{m}}{f^{\prime\prime}} -\frac13\frac{\mu^{m}}{f^{\prime\prime}} -\frac13 \frac{f^{\prime}}{f^{\prime\prime}}R + \frac23\frac{f}{f^{\prime\prime}}  
- \frac{f^{\prime\prime\prime}}{f^{\prime\prime}}X^{2}- X( \phi + \A ).\label{eq4}
\ee
When $f(R)=R$, we recover the GR description of the field equations, fluid quantities, and propagation equations.
\section{The TOV equations in the (1+1+2) covariant formalism}\label{sec:tov}
We will now derive the key equations that describe a compact stellar object in the context of $f(R)$ in the language of the covariant formalism summarized above. These equations will be equivalent to the so-called TOV equations in \cite{Oppenheimer:1939ne,Tolman:1939jz}. We will write them in terms of dimensionless variables, which will simplify the understanding of the mathematical structure of the equations and the resolution strategies we will employ.  

We start with the definition of a dimensionless radial parameter. We introduce the parameter, $\rho$, such that\
\be
\hat{X}=\phi X_{,\rho}.
\ee
To aid in the physical interpretation of our results, we can connect the parameter $\rho$ to the area radius $r$,
\be
\rho=2\ln\left(\frac{r}{r_{0}}\right),
\ee
where $r_{0}$ is an integration constant and it is set to $r_{0}=1$. In the following, we will use $\rho$ for the calculations, but the results will be reported in terms of $r$ so that it connects more easily with the existing literature.

Next, we introduce the following  normalized variables:
\bea
\Xi&=&\frac{\phi_{,\rho}}{\phi}\;,\; \quad Y=\frac{\A}{\phi}\;,\label{toVvar01}\\
\frac{X}{\phi}&\equiv&\mathbb{X}\;,\; \quad \mathcal{K}=\frac{K}{\phi^{2}}\;,\; \quad E = \frac{\varepsilon}{\phi^{2}},\label{toVvar1}\\
\tilde{\mathbb{M}}^{m}&=&\frac{\tilde{\mu}^{m}}{\phi^{2}}\;,\; \quad \tilde{P}^{m}=\frac{\tilde{p}^{m}}{\phi^{2}}\;,\; \quad \tilde{\mathbb{P}}^{m}=\frac{\tilde{\Pi}^{m}}{\phi^{2}},\label{toVvar2}\\
\mathbb{M}^{R}&=&\frac{\mu^{R}}{\phi^{2}}\;,\; \quad P^{R}=\frac{p^{R}}{\phi^{2}}\;,\; \quad \mathbb{P}^{R} = \frac{\Pi^{R}}{\phi^{2}}.\label{toVvar3}
\eea
Employing the general equations (\ref{equation1a}-\ref{equation5a}), the TOV equations for a general $f(R)$ gravity model with a baryonic matter source in the (1+1+2) covariant formalism read
\bea
&&\mathbb{X}_{,\rho}+\mathbb{X}\Xi = \frac{P^{m}}{f^{\prime\prime}} - \frac{\mathbb{M}^{m}}{3f^{\prime\prime}} - \frac{f^{\prime}}{3f^{\prime\prime}}\frac{R}{\phi^{2}} + \frac{2}{3}\frac{f}{f^{\prime\prime}\phi^{2}}- \frac{f^{\prime\prime\prime}}{f^{\prime\prime}}\mathbb{X} - \mathbb{X}(1+Y),\\
&&P^{\text{tot}}_{,\rho} + \mathbb{P}^{\text{tot}}_{,\rho} = - Y\left(\mathbb{M}^{\text{tot}} + P^{\text{tot}}\right)- \mathbb{P}^{\text{tot}}\left(2\Xi + Y + \frac{3}{2}\right)- 2\Xi P^{\text{tot}},\label{Pm1}\\
&&Y_{,\rho} = -Y\left(\Xi + Y + 1 \right)+\frac{1}{2}\left(\tilde{\mathbb{M}}^{m} + \mathbb{M}^{R}\right)+\frac{3}{2}\left(\tilde{P}^{m} + P^{R}\right),\label{Yevo}\\
&&\mathcal{K}_{,\rho} = -\mathcal{K}(1+2\Xi),\label{Kevo}
\eea
with the following constraints
\bea
1+4Y-4\mathcal{K} - 4(\tilde{P}^{m}+P^{R})-4(\tilde{\mathbb{P}}^{m} + \mathbb{P}^{R})&=& 0,\label{constraint1}\\
1 + 2\Xi - 2Y + 2(\tilde{\mathbb{M}}^{m} + \mathbb{M}^{R}) + 2(\tilde{P}^{m} + P^{R})+ 2(\tilde{\mathbb{P}}^{m} + \mathbb{P}^{R})&=& 0,\label{constraint2}\\
2(\tilde{\mathbb{M}}^{m} + \mathbb{M}^{R}) - 6Y - 6E 
+ 6(\tilde{P}^{m} + P^{R})+ 3(\tilde{\mathbb{P}}^{m} + \mathbb{P}^{R})
&=&0.\label{constraint3}
\eea

A general solution to the TOV equations may be given by the line element \cite{NCD}
\begin{equation}
	ds^{2} = -k_{1}(\rho)dt^{2} + k_{2}(\rho)d\rho^{2} + k_{3}(\rho)d\Omega^{2},\label{generic metric}
\end{equation}
where
\bea
k_{3}(\rho)&=&K_{0}e^{\rho},\\
d\Omega^{2}&=&d\theta^{2}+\sin^{2}\theta d\phi^{2},
\eea
and $K_{0}$ is a constant. The variables describing a static LRS-II spacetime in terms of the metric in \eqref{generic metric} and the parameter $\rho$ are
\bea
\phi&=&\frac{1}{\sqrt{k_{2}}}\;,\; \quad Y=\frac{k_{1,\rho}}{2k_{1}}\;,\label{covcoord1}\\
\Xi&=&-\frac{k_{2,\rho}}{k_{2}}\;,\; \quad \mathcal{A}=\frac{k_{1,\rho}}{2k_{1}\sqrt{k_{2}}}\;,\label{covcoord2}\\
\mathcal{K}&=&\frac{k_{2}}{K_{0}e^{\rho}}\;.\label{covcoord3}
\eea
The metric coefficients of \eqref{generic metric} are written in terms of  the area radius, $r$, as
\be
k_{1}(\rho)=k_{1}(r), ~~~~
k_{2}(\rho)=\frac{r^{2}}{4}k_{2}(r), ~~~~
r^{2}(\rho)=K_{0}e^{\rho}.
\ee

To find realistic solutions to the TOV equations, we will need to define and impose the physical and boundary conditions of our two-fluid compact stellar object. We address this in the next section.

\section{Physical and Boundary Conditions} \label{sec:physbound}
Not all solutions to the TOV equations represent physical relativistic stars. In fact, majority of the TOV solutions cannot correspond to any meaningful matter spacetime configuration. 
Despite this drawback, we can still define some minimum conditions that can be used to recognize more realistic solutions. To aid in this task, we define two additional thermodynamical potentials: radial pressure and tangential pressure, which are defined as  
\be
p_{r} = p+\Pi, ~~~~
p_{\perp} = p - \frac{1}{2}\Pi.\label{def_tang_pressure}
\ee
With these definitions, we can formulate the two types of constraints needed to describe a realistic relativistic compact object: thermodynamical constraints and junction conditions.

\subsection{Thermodynamical constraints}\label{sec:fluid constraints}
We start with the constraint on the thermodynamical quantities. A solution to the TOV equations can represent a physical relativistic star if the energy density, radial pressure, and the tangential pressure are positive inside the star, i.e.,
\be
\mu^{m}\geq0, ~~~~ p^{m}_{r}\geq 0, ~~~~ p^{m}_{\perp}\geq0.\label{Eq: matter source constraints}
\ee
In GR it is often also required that the gradients of these quantities are negative within the relativistic star. However, as we shall see, this is not necessarily true in our context. The conditions above imply that the weak energy condition, 
\be
\mu^{m}+p^{m}_{r}\geq0,
\ee
is always satisfied.
The speed of sound of the matter sources has to obey the causal limits:
\be
0\leq c_{m, r}^{2}=\frac{\partial p^{m}_{, r}}{\partial\mu^{m}}\leq1, ~~~~~ 0\leq c_{m, \perp}^{2}=\frac{\partial p^{m}_{, \perp}}{\partial \mu^{m}}\leq1,
\ee
so that no sound wave can travel faster than the speed of light.

Note that the above conditions apply only to the standard matter quantities. The curvature fluid and the effective fluid associated with matter in $f(R)$ gravity can violate these conditions without compromising the physical interpretation of the solutions.

\subsection{Junction conditions}\label{sec:junction conditions}
It is customary in relativistic astrophysics to assume that compact stellar objects have a ``hard'' boundary, i.e., matter is confined in a well-defined volume surrounded by a vacuum. The most convenient way to describe this configuration is to simply join the interior spacetime with a vacuum exterior spacetime. A set of general, covariant conditions that allow joining two different spacetimes are due to Israel \cite{Israel}. Assuming, as in our case, that the normal $n_a$  of the boundary coincides with $e_a$, the junction conditions read as:
\begin{align}
[\gamma_{ab}]^+_-&=0\,,\label{JCGR1}
\\
[K_{ab}]^+_- - \gamma_{ab} [K]^+_-&=- S_{ab}\,,\label{JCGR2}
\end{align}
where $\gamma_{ab}=N_{ab}+u_{a}u_{b}$ is the induced metric on the separation surface, $K_{ab}$ is the extrinsic curvature, $S_{ab}$ represents the stress-energy tensor of a possible shell within the boundary surface $\mathcal{S}$. We have employed the notation $ [\chi]^+_-=\chi^+-\chi^-$ which, for simplicity, will be denoted as {\it jump of $\chi$}. For later convenience, we also define 
\be
\{\chi\}=\frac{1}{2}(\chi^{+}+\chi^{-}).
\ee
The above conditions, which are purely geometric, can be converted into simple conditions on the baryonic matter's thermodynamical potentials.  In particular, using the Einstein field equations, one obtains that 
\be
S_{ab}\{K^{ab}\}+[T_{ab}e^a e^b]^+_-=0,
\ee
which in the case of the soldering of static spherically symmetric metrics, implies 
\be
[p_r]^+_-=0.
\ee

In the case of $f(R)$ gravity, the Israel junction conditions must be extended to account for the additional degree of freedom carried by the higher-order terms. These conditions were first presented in \cite{Deruelle:2007pt} and successively expanded in \cite{Senovilla:2013,Senovilla:2014}, where some peculiar aspects of the junction in these theories are presented\footnote{See also \cite{LuisRosa:gen-mod-grav-junctions} for a general review on junction conditions for modified theories of gravity}. In four dimensions, we have
\begin{align}
[\gamma_{ab}]^+_-&=0 \label{JCfR1}\,,
\\
[K]^+_-&=0\,,\label{JCfR2}
\\
[R]^+_-&=0\,,\label{JCfR3}
\\
f'(R)[K^*_{ab}]^+_- &= -S^*_{ab}\,,\label{JCfR4}
\\
3f''(R)[e^a\nabla_{a}R]^+_-&= S  \label{JCfR5}\,,
\end{align}
where
\begin{align}
K^*_{ab}&=K_{ab}-\frac{1}{3}\gamma_{ab} K ,
\\
S^*_{ab}&=S_{ab}-\frac{1}{3} \gamma_{ab} S.
\end{align}
As we are adopting the effective fluid perspective and in line with what is usually done in GR, it will be useful to translate the above equations into constraints on the effective thermodynamical quantities. For the case of a two-fluid system in GR, which is equivalent to our case, the Israel conditions amount to
 \be
S_{ab}\{K^{ab}\}+[T^{\text{tot}}_{ab}e^ae^b]^+_-=0,
\ee
which, in our case and supposing the absence of a shell, implies 
\be
 [T^{\text{eff}}_{ab}e^ae^b]^+_-=[p^{\text{eff}}_r]^+_-= 0,
\ee
and therefore, form the definition of $p^{\text{eff}}_r$,
\be
 \left[\frac{p^{m}_r}{f^{\prime}}+p^{R}_r\right]^+_-= \left[\frac{p^{m}_r}{f^{\prime}}\right]^+_- +\left[p^{R}_r\right]^+_-=0.
\ee 
Assuming that the function $f$ does not contain a different cosmological constant term in the interior and exterior,  Eq. \eqref{JCfR3} implies that the jump of $f$ and its derivatives with respect to $R$ are zero. As a consequence, we can write
\begin{align}
   [p^{R}_{r}]^+_-&=\left[p^{R}+ \Pi ^{R}\right]^+_-\nonumber
   \\
   &=\left[\frac{f}{2f^{\prime}}\right]^+_- - \left[\frac{f''}{f'}\right]^+_- \{X\}\{\phi\} - \left[\frac{f''}{f'}\right]^+_- \{X\}\{\mathcal{A}\},
\end{align}
where we have used the properties 
\begin{align}
[a+b]^+_- =& [a]^+_- + [b]^+_-,\\
[a\cdot b]^+_- = & \{a\}[b]^+_- + \{b\}[a]^+_-\nonumber\\
= & \frac{1}{2}(a^{+} + a^-)(b^{+} - b^-) + \frac{1}{2}(b^{+} + b^-)(a^{+} - a^-).
\end{align}
Hence, we can conclude that  Eqs. (\ref{JCfR1}-\ref{JCfR5}), imply $ [p^{R}_{r}]^+_-=0$ and that a smooth junction requires
\begin{equation}\label{jcpr}
\left[p^m_r\right]^+_-= \left[p^m+\Pi^m\right]^+_-=0.   
\end{equation}
 This is consistent with the results of \cite{Senovilla:2013,Senovilla:2014}. However, as opposed to GR, this is not the only condition on the matter thermodynamics. In fact, since  the gravitational field equations can be written as
\begin{equation}
R=3 p^{\text{tot}}-\mu^{\text{tot}}.
\end{equation}
Equation \eqref{JCfR3} implies 
\bea
    0=\left[R\right]^+_-
    &=&\left[\frac{3 p^{m}-\mu^{m}}{f'} -\frac{f''}{f'}\hat X\right]^+_-\nonumber\\
    &=&\left[3 p^{m}-\mu^{m}\right]^+_- -\frac{f''}{f'}\left[\hat X\right]^+_-,
    \label{ricci-bc}
\eea
must hold at the boundary. As a result, this relation implies a constraint on the energy density and isotropic pressure at the boundary. 

The junction conditions mentioned above indicate that one can compensate for a mismatch in the extrinsic curvature or in the derivative of the Ricci scalar along the normal by assuming the boundary $\mathcal{S}$ is represented by a specific matter distribution given by the tensor $S_{ab}$. We can calculate the components of this tensor by recognizing that the extrinsic curvature of $\mathcal{S}$ is 
\begin{equation}
\begin{split}
    K_{ab} &= \gamma_{a}{}^{c}\gamma_{b}{}^{d}\nabla_{c}e_{d}\\
    &=(N_{a}{}^{c} + u_{a}u^{c})(N_{b}{}^{d} + u_{b}u^{d})\nabla_{c}e_{d}.
    \end{split}
\end{equation}
In the spherically symmetric case, the jump of the extrinsic curvature is then given by
\begin{equation}
    [K_{ab}]^{+}_{-}=\left[\frac{1}{2}\phi N_{ab} - u_{a}u_{b}\mathcal{A}\right]^{+}_{-}.
\end{equation}
Then Eq. \eqref{JCfR4} and Eq. \eqref{JCfR5} imply that the stress-energy tensor on the boundary is given by
\begin{align}
    S_{ab} &=(N_{ab}+u_{a}u_{b})  f'' [X]^{+}_{-} -  f'[K_{ab}]^{+}_{-} \nonumber\\
    &= \left(f' [\mathcal{A}]^{+}_{-}+ f''[X]^{+}_{-}\right)u_{a}u_{b} + \left(f''[X]^{+}_{-} - \frac{f'}{2}\left[ \phi\right]^{+}_{-}\right)N_{ab}. \label{eq: SET boundary}
\end{align}
The shell will have energy density and orthogonal pressure
\begin{align}
&\mu^{\mathcal{S}}=S_{ab}u^{a}u^{b},\label{eq: pre stable fluid shell 1}\\
&p^{\mathcal{S}}_\perp =\frac{1}{2}S_{ab}N^{ab}.\label{eq: pre stable fluid shell 2}
\end{align}
In this case the standard requirement for $\mu^{\mathcal{S}}$ and $p^{\mathcal{S}}_\perp$  is to be non negative. However, one can still consider negative values of this last quantity taking into account that the shell matter still satisfy the weak energy condition. If this is the case then the condition $p^{\mathcal{S}}_{\perp}<0$ simply implies that the shell matter presents a tension. Notice that the radial pressure at the surface $\mathcal{S}$ is zero, i.e., $p^{\mathcal{S}}_{r}=S_{ab}e^{a}e^{b}=0$, as expected.

Finally,  it was shown in \cite{Senovilla:2013,Senovilla:2014} that shells in $f(R)$-gravity can have a more complex stress-energy tensor than the shells in GR. Although not immediately clear from the junction conditions, Eqs. (\ref{JCfR1}-\ref{JCfR5}),  these shells can present a  so-called {\it double layer}. In the context of $f(R)$ theories, structures of this kind can appear when the condition Eq. \eqref{JCfR3} is violated in theories where $f'''(R)=0$.
For these theories, the stress-energy tensor on the boundary acquires several additional components along the normal, which are related to the value of $\left[R\right]^+_-$. Indeed, the total stress-energy tensor of the shell will be given by
\begin{align}
    \bar{S}_{ab} +\bar{\varsigma}_{ab}&=S_{ab}+\varsigma_{ab}+2\varsigma_{(a} e_{b)}+ \varsigma e_a e_b+\bar{\varsigma}_{ab},\label{eq: pre tot double layer EMT}
\end{align}
where
\begin{align}
    \varsigma_{ab}&= f'' \left\{K_{ab}\right\} \left[R\right]^+_-,\\
     \varsigma_{a}&= f'' \left(N^{b}{}_{a}+u^{b}u_{a}\right) \nabla_b \left[R\right]^+_-,\\
     \varsigma&= f'' \left\{K\right\} \left[R\right]^+_-,
\end{align}
and $\bar{\varsigma}_{ab}$ represents  the energy-momentum content of the double layer. This is akin to a dipole distribution, and it is given by
\be
\bar{\varsigma}_{ab}=f'' \nabla_\rho \left[\left[R\right]^+_- \gamma_{ab} e^\rho \delta\right]=f''\Delta_{ab},
\ee
where $\delta$ represents Dirac's delta, $\Delta_{ab}$ is the double layer distribution and $f''$ is a constant.

Notice that the presence of $\varsigma_{a}$ and $\varsigma$ also requires the presence of $\bar{\varsigma}_{ab}$, but the converse is not necessarily true.

Decomposing $\bar{S}_{ab}$ along $u^a$, $e^a$ and $N_{ab}$ leads to
\begin{align}
    \bar{S}_{ab}&=\bar{\mu}^S u_a u_b + \bar{p}_r^S e_a e_b + \bar{p}_\perp^S N_{ab} +2 \bar{Q}^S u_{(a}e_{b)}+\bar{Q}^S_{(a}e_{b)},\label{eq: tot double layer EMT}
\end{align}
where
\begin{align}
&\bar{\mu}^{\mathcal{S}}=f' [\mathcal{A}]^{+}_{-}+ f''[X]^{+}_{-}-f''\{\mathcal{A}\}[R]^{+}_{-},\label{eq:stable fluid shell 1}
\\
&\bar{p}_r^{\mathcal{S}} =f''\{K\}[R]^{+}_{-},\label{eq:stable fluid shell 2}
\\
&\bar{p}_\perp^{\mathcal{S}} = - \frac{1}{2}f'\left[ \phi\right]^{+}_{-}
+f''[X]^{+}_{-} + f''\{\phi\}[R]^{+}_{-},\label{eq:stable fluid shell 3}
\\
&\bar{Q}^S = f'' \left(u^b\nabla_b [R]^{+}_{-}\right)u_{a},\label{eq:stable fluid shell 4}
\\
&\bar{Q}_a^S = f''\delta_{a}[R]^{+}_{-}.\label{eq:stable fluid shell 5}
\end{align}
Instead for $\bar{\varsigma}_{ab}$   we can write
\begin{equation}
    \bar{\varsigma}_{ab}=f'' \left(\Delta_u u_a u_b+\frac{1}{2} \Delta_N N_{ab}  \right),\label{eq: double layer EMT dec}
\end{equation}
where $\Delta_u=\Delta_{ab}u^a u^b$ and $\Delta_N=\Delta_{ab}N^{a b}$.
\section{Reconstruction of exact solutions} \label{sec: solutions method}
In this section, on the basis of the results obtained in \cite{NCD,NCD2} and using the $f(R)$ TOV equations, Eqs. (\ref{Pm1}-\ref{constraint3}), we will develop a reconstruction technique that will allow us to generate several exact solutions describing compact stellar objects.
\\
We start by assuming a form for the metric tensor. This determines the quantities $Y$ and $\mathcal{K}$, as they are related to the metric coefficients with Eqs. \eqref{covcoord1}--\eqref{covcoord3}, and we can compute the Ricci scalar in terms of these quantities as well. By specifying our $f(R)$ function, we can determine the thermodynamical description of our curvature fluid in terms of the metric coefficients.
\\
If we consider Eqs. \eqref{Yevo}, \eqref{Kevo}, \eqref{constraint1} and \eqref{constraint2}, we can find new solutions to the matter fluid from quantities that are constructed from the metric alone:
\bea
&&R = \frac{\phi^2 \left(\mathcal{K} \left(4 \mathcal{K}-4 Y_{,\rho}-2 Y (2 Y+1)-1\right)+2 (Y+1) \mathcal{K}_{,\rho}\right)}{2 \mathcal{K}},\label{ricci-coord-expr}\\
&&\widetilde{\mathbb{M}}^{m} = -\frac{-2 \mathcal{K}_{,\rho}-4 \mathcal{K}^2+\mathcal{K}+4 \mathcal{K} \mathbb{M}_{R}}{4 \mathcal{K}},\label{Mm-sol}\\
&&\widetilde{P}^{m} = -\frac{2 \mathcal{K}_{,\rho}+4 \mathcal{K}^2-\mathcal{K}+12 \mathcal{K} P_{R}-8 \mathcal{K} Y_{,\rho}+4 Y \mathcal{K}_{,\rho}-8 \mathcal{K} Y^2-4 \mathcal{K} Y}{12 \mathcal{K}},\label{Pm-sol}\\
&&\widetilde{\mathbb{P}}^{m} = -\frac{6 \mathbb{P}_{R} \mathcal{K}-\mathcal{K}_{,\rho} + 4 \mathcal{K}^2 - \mathcal{K} + 4 \mathcal{K} Y_{,\rho} - 2 Y \mathcal{K}_{,\rho} + 4 \mathcal{K} Y^2 - 4 \mathcal{K} Y}{6 \mathcal{K}},\label{aniPm-sol}
\eea
where
\begin{align}
    \mathbb{M}^{R} &= \frac{R}{2\phi^{2}} + \frac{f^{\prime\prime}}{f^{\prime}}\left(R_{,\rho\rho} + R_{,\rho} + \Xi R_{,\rho}\right) - \frac{1}{2\phi^{2}}\frac{f}{f^{\prime}},\\
    P^{R} &= - \frac{R}{2\phi^{2}} - \frac{2}{3}\frac{f^{\prime\prime}}{f^{\prime}}\left(R_{,\rho\rho} + R_{,\rho} + \Xi R_{,\rho} + \frac{3}{2}YR_{,\rho}\right)+\frac{1}{2\phi^{2}}\frac{f}{f^{\prime}},\\
    \mathbb{P}^{R} &= \frac{2}{3}\frac{f^{\prime\prime}}{f^{\prime}}\left(R_{,\rho\rho} - \frac{1}{2}R_{,\rho} + \Xi R_{,\rho}\right).
\end{align}
Equations (\ref{Mm-sol}--\ref{aniPm-sol}) satisfy the constraints, Eqs. (\ref{constraint1}--\ref{constraint3}), and therefore, naturally satisfy the TOV equations. Although Eqs. (\ref{Mm-sol}--\ref{aniPm-sol}) would represent an infinite number of solutions, not all of these solutions have physical value. More specifically, it is imperative that the boundary (junction) and physical conditions, discussed in Sec. \ref{sec:physbound}, are satisfied in order to describe realistic relativistic stars in the context of $f(R)$ gravity. 

The reconstruction approach described above requires choosing a specific form for the function $f$. In the following, we consider the quadratic model \eqref{Star action} mentioned in Sec. \ref{sec:FE}. Our choice is motivated by three considerations: (i) its relevance in cosmology and quantum field theory in curved spacetime, (ii) its simplicity, which, as we will see, will be an important issue in the derivation of exact solutions, and (iii) the fact that this model is the simplest $f(R)$ gravity model that allows the exploration of double layers. 

Once $f$ is fixed, we need to choose the base metric for the reconstruction algorithm. The obvious initial choice is to consider the metric coefficients of two well-known single fluid solutions in GR and combine them to obtain a new solution. The advantage of this approach is that it is more likely to obtain physically meaningful solutions, including the fact that the mismatch will naturally generate the anisotropic pressure needed in the fluid representation of $f(R)$ gravity. However, this choice is not always the most convenient, and for this reason, we will also consider a completely general starting metric.

\section{Resconstruction of quadratic models with Interior Schwarzschild-Tolman IV metric} \label{sec:S-T}

Let us combine two of the simplest descriptions of the interior of a relativistic star: the Interior Schwarzschild and the Tolman IV solutions. In particular, we choose the  component $k_{1}$ of the  solution metric corresponding to the one of  interior Schwarzschild metric and the component $k_{2}$ as corresponding to the Tolman IV solution in terms of the area radius
\be
k_{1}(r)=a_{0}(c_{1}+z)^{2}, ~~~~~
k_{2}(r)=\frac{\mathcal{R}^{2}(A^{2}+2r^{2})}{(\mathcal{R}^{2}-r^{2})(A^{2}+r^{2})},\label{eq: metric coeffs 1 const density tolman 4}
\ee
where $z=\sqrt{3-\mu_{1}r^{2}}$ and $a_{0}$, $c_{1}$,  $\mu_{1}$, $A$ and $R$ are constants. Note that $\mu_{1}$ is a constant in the original solution and is related to the (constant) density of the source. However, in this context, it is simply an additional parameter.

In terms of the parameter $\rho$ and the variables (\ref{toVvar01}-\ref{toVvar3}), these correspond to
\bea
Y_{\text{IS}} &=& \frac{\mu_{1}e^{\rho}}{2\mu_{1}e^{\rho}-2c_{1}\sqrt{3-\mu_{1}e^{\rho}}-6},\label{Y-int-schwarz}\\
\mathcal{K}_{\text{TIV}} &=& \frac{-\mathcal{R}^{2}(A^{2}+2e^{\rho})}{4(A^{2}+e^{\rho})(e^{\rho}-\mathcal{R}^{2})},\label{K-TIV}\\
\phi &=& \frac{e^{-\frac{\rho}{2}}}{\sqrt{\mathcal{K}_{\text{TIV}}}}.
\eea

As mentioned, the real challenge in finding physical solutions is that strict boundary and thermodynamical conditions must be satisfied. In particular, we will need to ensure that sets of parameters exist for which the conditions outlined in Sec. \ref{sec:fluid constraints} are satisfied. In addition, we choose to match this solution with an exterior Schwarzschild solution. With the anzats, Eqs. (\ref{Y-int-schwarz}-\ref{K-TIV}), the expression for the Ricci scalar $R$, Eq. \eqref{ricci-coord-expr} is independent of the Starobinsky parameter $\alpha$, and we could only find a solution that satisfies the physical constraints when $\alpha=0.001$. In natural units, this corresponds to $\alpha\sim10^{7}~cm^{2}$ which is compatible with the constraint found in \cite{ConstraintsonfRviaNeutronStars}.

\begin{figure}[h]
  \centering
  \includegraphics[scale=0.6]{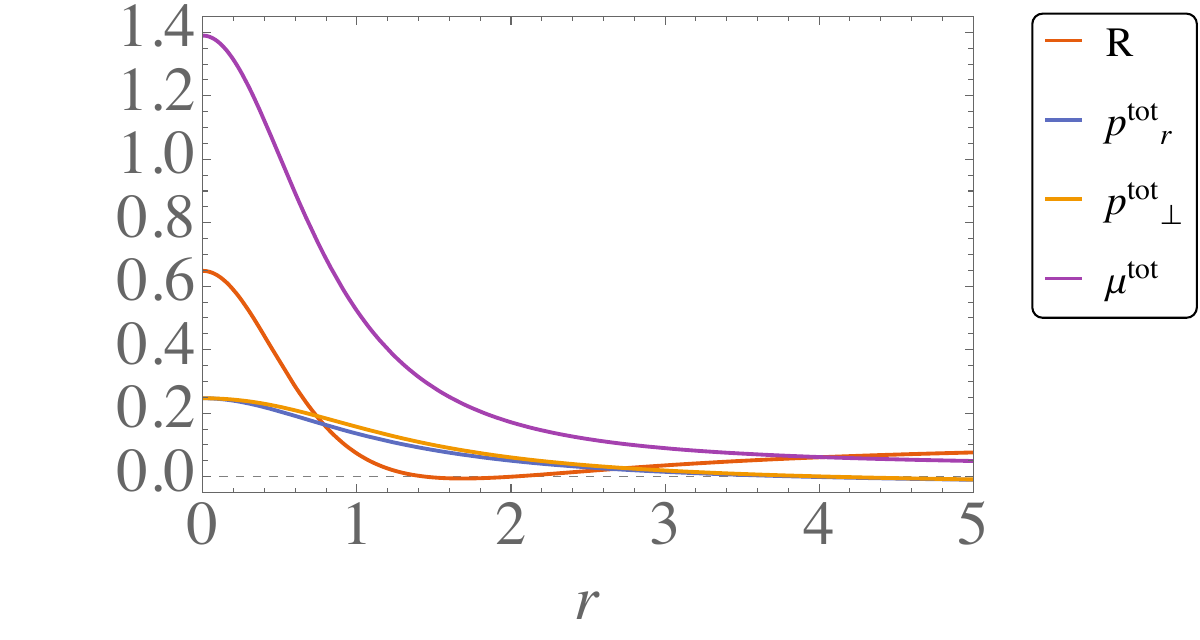}
  \caption{Solutions to the quadratic $f(R)$ model with $\alpha=0.001$ for the interior Schwarzschild-Tolman IV (IS-TIV) geometry in Sec. \ref{sec:S-T}. The parameter values are: $\mu_{1}=-1.25$, $\mathcal{R}=7.3$, $c_{1}=0.3$, and $A=1.5$. Here, $r$ is the normalized area radius, i.e. $r/r_{0}$ with $r_{0}=1$.}
  \label{fig:starobinsky soln 1}
\end{figure}

\begin{figure}
    \centering
    \includegraphics[scale=0.6]{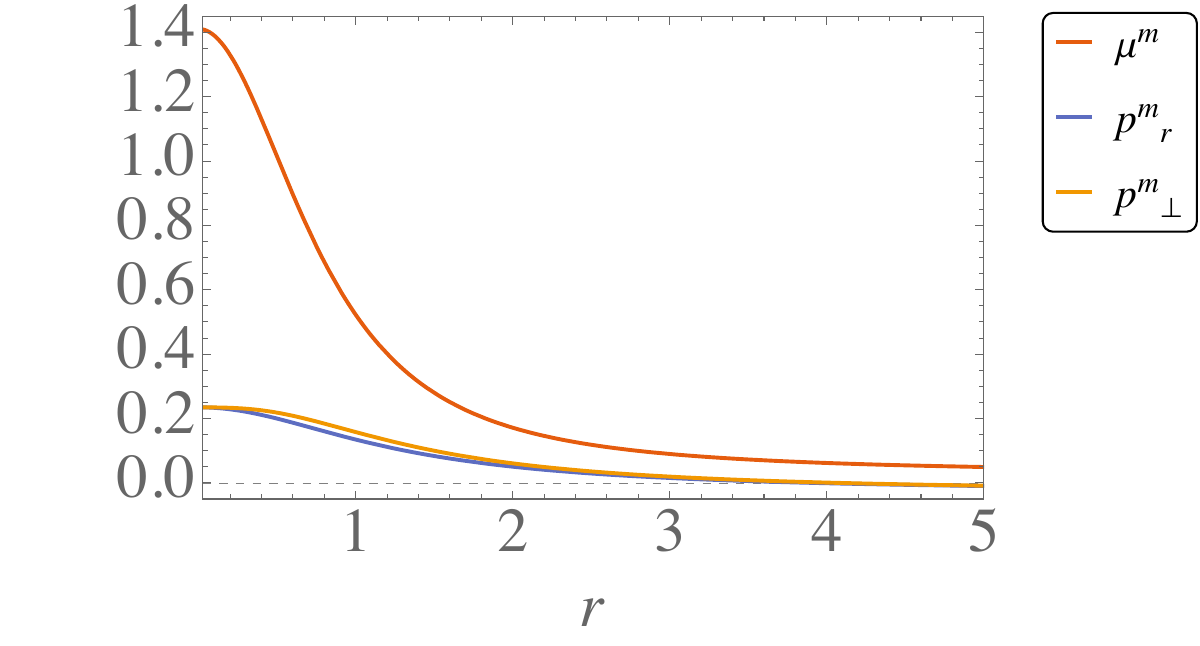}
    \caption{The matter fluid solutions for the (IS-TIV) geometry in Sec. \ref{sec:S-T}. The energy density, radial and tangential pressure satisfy the physical conditions in Sec. \ref{sec:physbound}. The parameter values are: $\alpha=0.001$, $\mu_{1}=-1.25$, $\mathcal{R}=7.3$, $c_{1}=0.3$, and $A=1.5$. Here, $r$ is the normalized area radius, i.e. $r/r_{0}$ with $r_{0}=1$.}
    \label{fig: ricci-scalar-tot-pressure}
\end{figure}

\begin{figure}
    \centering
    \includegraphics[scale=0.6]{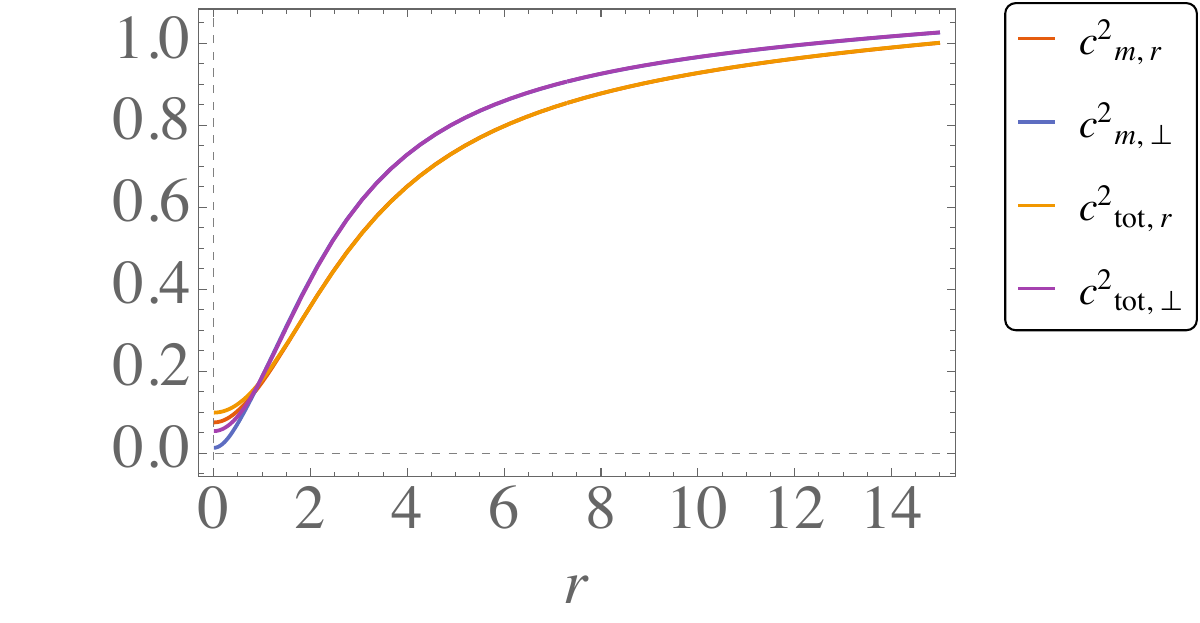}
    \caption{The radial and tangential speed of sound of the matter fluid, and the speed of sound for the total radial and orthogonal fluid quantities for the (IS-TIV) geometry in Sec. \ref{sec:S-T}. The causal conditions in Sec. \ref{sec:physbound} are satisfied. The parameter values are: $\alpha=0.001$, $\mu_{1}=-1.25$, $\mathcal{R}=7.3$, $c_{1}=0.3$, and $A=1.5$. Here, $r$ is the normalized area radius, i.e. $r/r_{0}$ with $r_{0}=1$.}
    \label{fig:starobinsky soln 44}
\end{figure}

\begin{figure}
    \centering
    \includegraphics[scale=0.6]{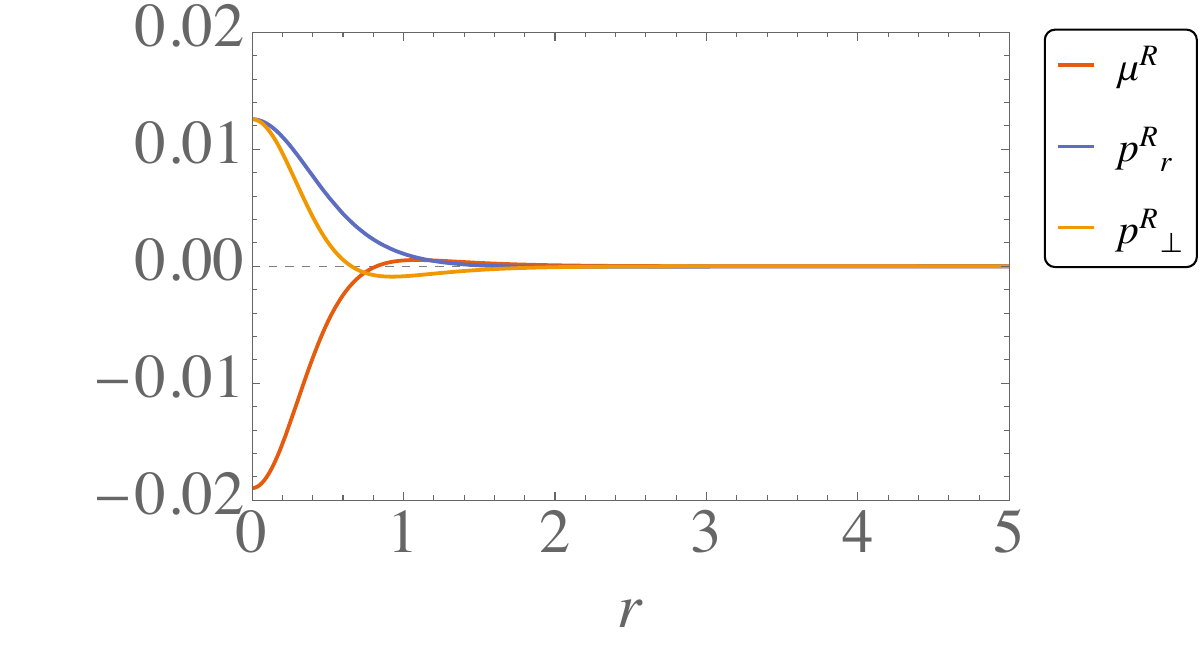}
    \caption{The curvature fluid solutions for the (IS-TIV) geometry in Sec. \ref{sec:S-T}. The parameter values are: $\alpha=0.001$, $\mu_{1}=-1.25$, $\mathcal{R}=7.3$, $c_{1}=0.3$, and $A=1.5$. Here, $r$ is the normalized area radius, i.e. $r/r_{0}$ with $r_{0}=1$.}
    \label{fig:starobinsky soln 3}
\end{figure}

\begin{figure}
\centering
\begin{subfigure}{\textwidth}
  \centering
  \includegraphics[scale=0.8]{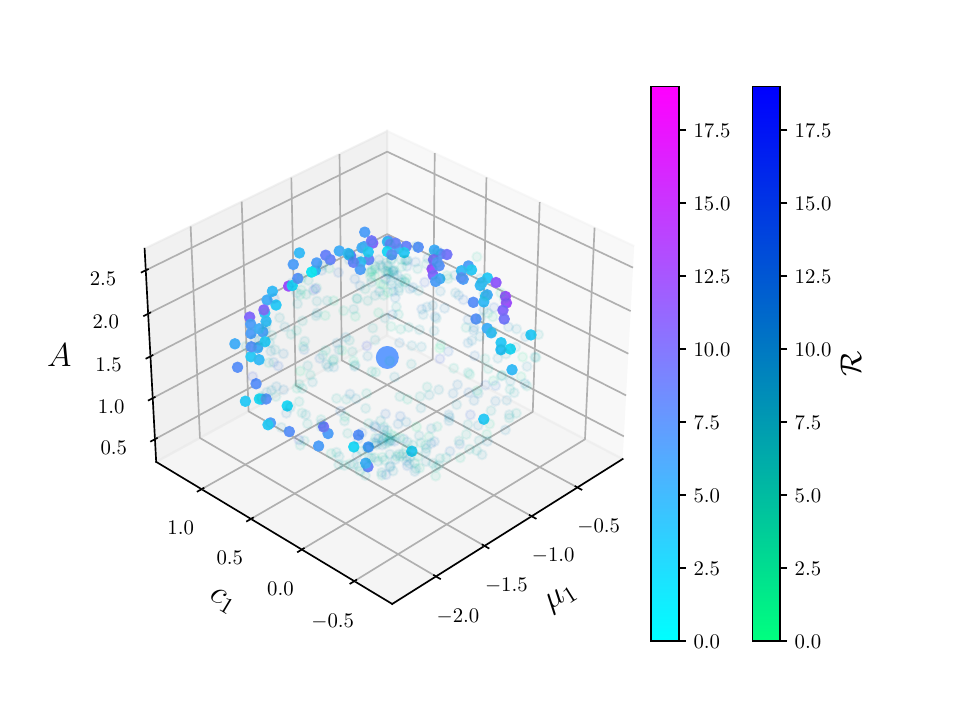}
  \caption{Parameter space plot for the radial, squared speed of sound for the baryonic matter fluid.}
  \label{fig:starobinsky soln 5}
\end{subfigure}%

\begin{subfigure}{\textwidth}
  \centering
  \includegraphics[scale=0.38]{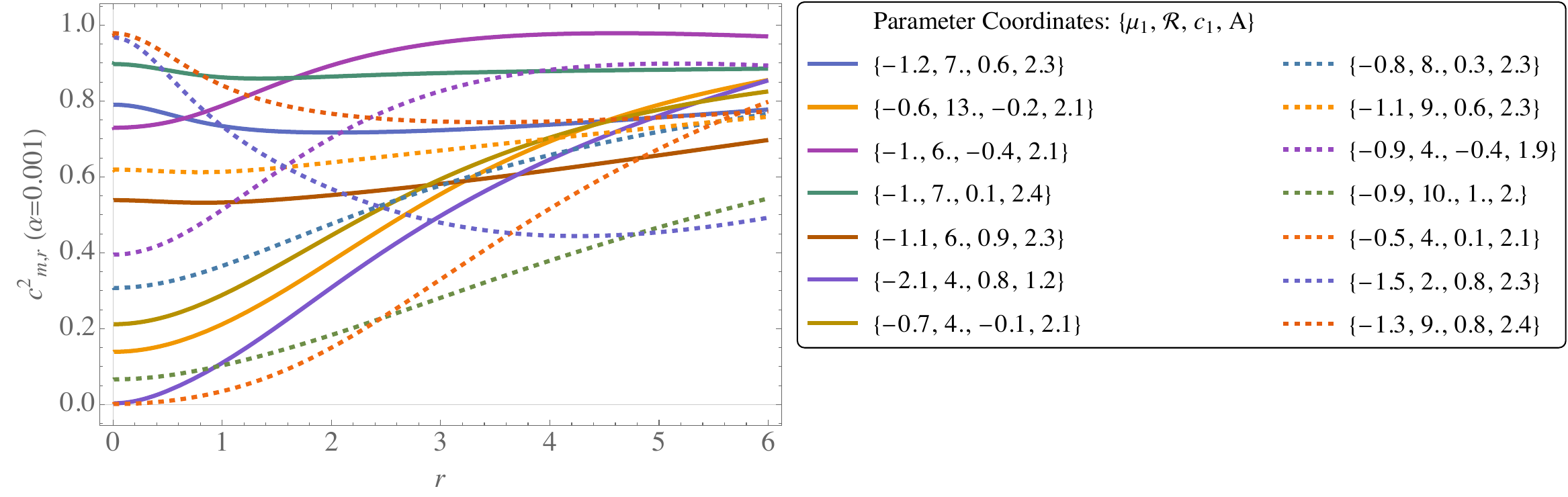}
  \caption{The radial, squared speed of sound for the baryonic matter fluid for the parameter values satisfying the causal condition in Fig. \ref{fig:starobinsky soln 5}. Here, $r$ is the normalized area radius, i.e. $r/r_{0}$ with $r_{0}=1$.}
  \label{fig:starobinsky soln 6}
\end{subfigure}
\caption{Figure \ref{fig:starobinsky soln 5} shows a perturbation in the parameter space from the central point which corresponds to the parameter values in Fig. \ref{fig:starobinsky soln 1}. The faint points are a generation of 500 random sets of parameter values with a radial shift of $0.05$ and constrained to a sphere of radius 1. The darker points away from the center, which are $21\%$ of the total points on the sphere, satisfy the causal condition $0<c_{m, r}^{2}\leq1$. This analysis is performed for the (IS-TIV) geometry in Sec. \ref{sec:S-T}. Figure \ref{fig:starobinsky soln 6} illustrates the general envelope of the solutions $c^{2}_{m,r}$, for the parameters that satisfy the causal conditions in Fig. \ref{fig:starobinsky soln 5}}
\label{fig: param space test}
\end{figure}

However, our solution with these chosen parameters presents a shell with a double layer. This is due to the fact that the Ricci scalar, $R$, and the matter radial pressure do not go simultaneously to zero at the boundary of the star.  We can then calculate the properties of the matter that compose the shell using the results of Sec. \ref{sec:physbound}. The total fluid thermodynamics on the surface $\mathcal{S}$, which includes  the shell and the double layer strength, in this case, are 
\begin{align}
    \bar{\mu}^{\mathcal{S}} &= 2\alpha[X]^{+}_{-}+(1+2\alpha R)[\mathcal{A}]^{+}_{-} - 2\alpha\{\mathcal{A}\}[R]^{+}_{-},\label{eq:density shell}\\
    \bar{p}^{\mathcal{S}}_{r} &= 2\alpha\left(\{\phi\}+\{\mathcal{A}\}\right)[R]^{+}_{-},\label{eq:pressure shell}\\
    \bar{p}^{\mathcal{S}}_{\perp} &= 2\alpha[X]^{+}_{-}-\frac{1}{2}(1+2\alpha R)[\phi]^{+}_{-} + f''\{\phi\}[R]^{+}_{-},\label{eq:ptbar}
\end{align}
Using the parameter values in Figs. \ref{fig:starobinsky soln 1}--\ref{fig: param space test} the energy density along the surface $\mathcal{S}$ is $\bar{\mu}^{\mathcal{S}}>0$. However, the orthogonal pressure along the surface $\mathcal{S}$, is negative, i.e. $\bar{p}^{\mathcal{S}}_{\perp}<0$, while $\bar{p}^{\mathcal{S}}_{r}>0$. As, $\bar{\mu}^{\mathcal{S}}+\bar{p}^{\mathcal{S}}_{r}>0$  the weak energy condition is satisfied and we can conclude that the shell presents a tension within the boundary surface.
The full expressions of the jump quantities are in \ref{eq: full boundary quantities}.

Naturally, a single set of parameters that satisfies the physical requirements of Sec. \ref{sec:physbound} is not necessarily sufficient to validate the solution we have found. We also need to prove that sets of parameters exist for which the physical and boundary conditions are satisfied. In order to achieve this goal, we employed computational methods and, more specifically, a parameter space analysis that we will briefly discuss here.

We generated a list of 500 random numbers from a normal Gaussian distribution for each parameter constant in the interval $\{-10, 10\}$ for $\mu$, $A$ and $\mathcal{R}$ and $c_{1}$. Combinations of these lists are iterated through our analytical expressions for the radial pressures for the matter and curvature fluids for a fixed value of $\alpha$. 
Since the strictest physical constraint is the causal condition for the matter source, we implement a conditional statement that tests this condition for an iterated combination of the parameter values. We plot the combination of parameter values that satisfies this causal condition. Performing this routine, we look for regions of clustering of points in the parameter space. This narrows our parameter-space intervals and improves our chances of finding a solution that satisfies conditions \eqref{Eq: matter source constraints}. This methodological approach proved much more helpful in finding physical solutions than a trial-and-error approach. We present solutions in Figs. \ref{fig:starobinsky soln 1}--\ref{fig:starobinsky soln 44} where the physical conditions in Sec. \ref{sec:physbound} are satisfied.

As we have emphasized before, the curvature fluid is effective. Thus, its physical interpretation is not bounded by the constraints of baryonic matter. However, we can comment on its influence on the thermodynamics of the baryonic matter. We notice, immediately, that the energy density, radial and tangential pressures of the curvature fluid (cf. Fig. \ref{fig:starobinsky soln 3}) are small in comparison to the matter fluid solutions (cf. Fig. \ref{fig: ricci-scalar-tot-pressure}).  This effect, which in principle could be ascribed to the value of the parameter $\alpha$ (we have chosen to be $\alpha=0.001$), is not strictly related to it. In fact, in the proceeding section, we will deal with an even smaller value of this parameter that still leads to comparable thermodynamical potentials for standard matter and the curvature fluid. This should not be surprising as it is a consequence of the nonlinearity of the theory: small corrections to the Hilbert-Einstein action do not always lead to solutions close to GR ones.

\begin{figure}[ht]
\centering
  \includegraphics[scale=0.45]{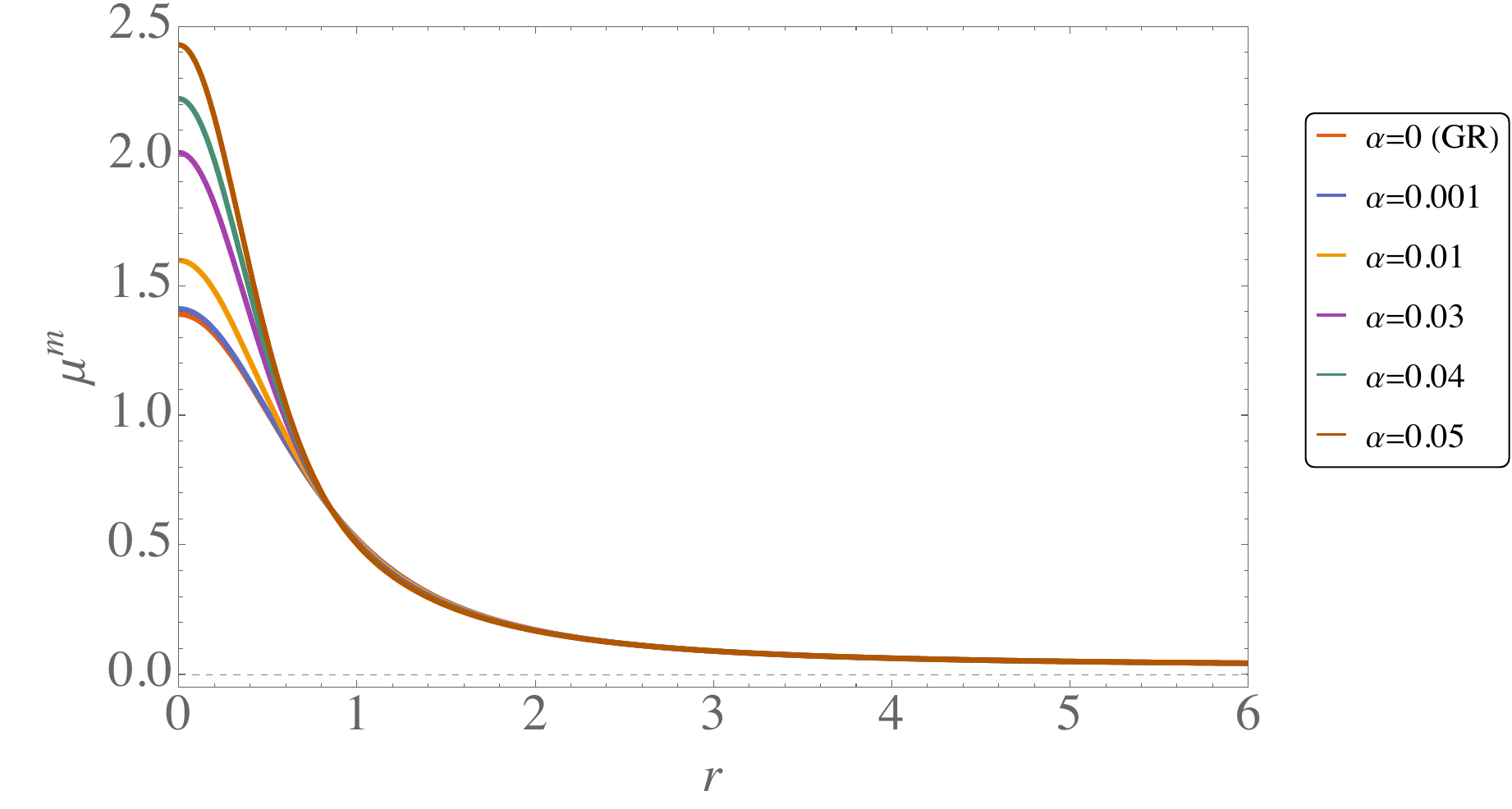}
  \caption{Energy density of the matter fluid for different values of $\alpha$, using the same parameter values as in Fig. \ref{fig:starobinsky soln 1}, for the (IS-TIV) geometry in Sec. \ref{sec:S-T}. Here, $r$ is the normalized area radius, i.e. $r/r_{0}$ with $r_{0}=1$.}
  \label{fig:matter vary alpha 1}
\end{figure}

\begin{figure}[ht]
\centering
\begin{subfigure}{\textwidth}
  \centering
  \includegraphics[scale=0.45]{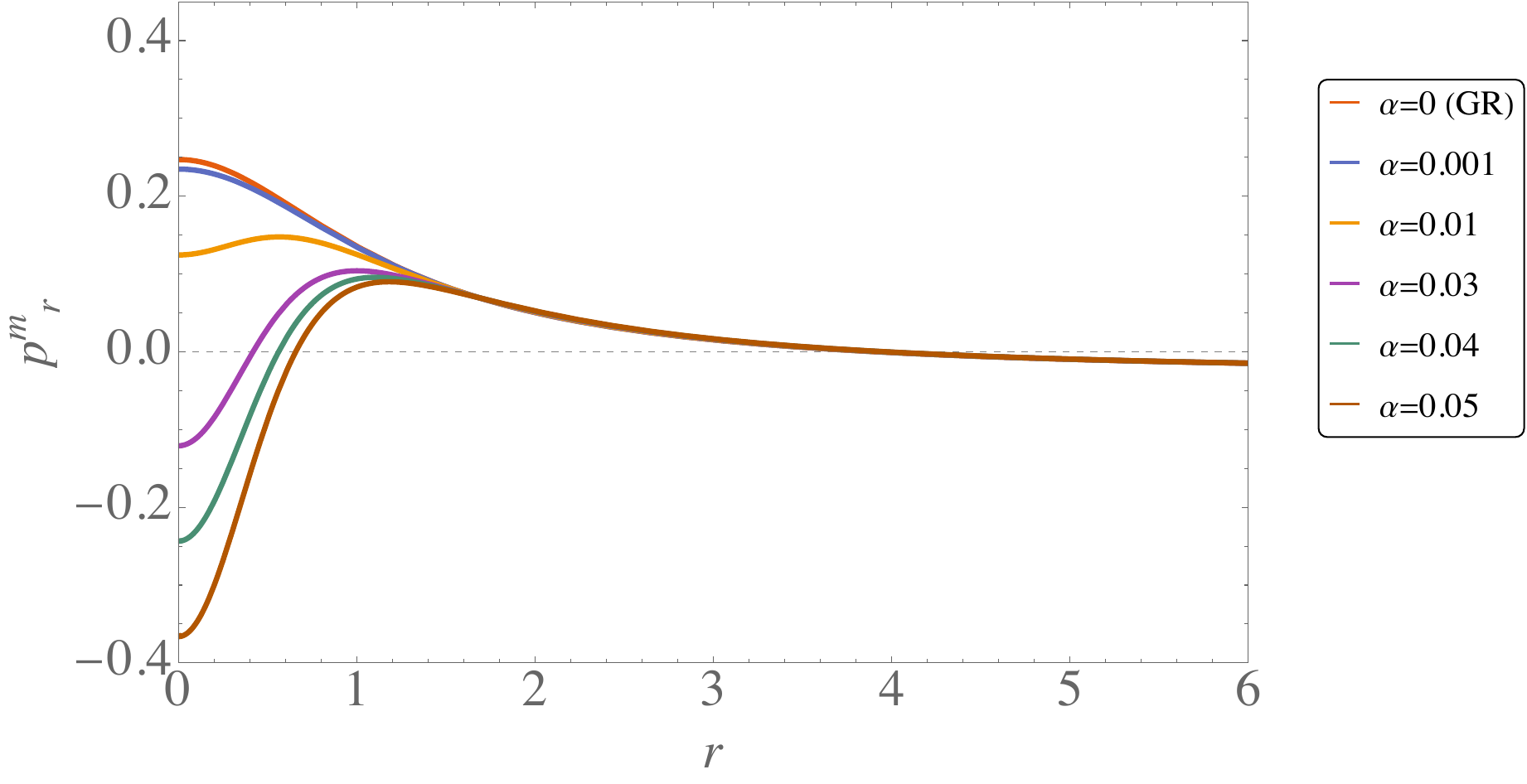}
  \caption{Radial pressure of the matter fluid for different values of $\alpha$ when considering the (IS-TIV) geometry in Sec. \ref{sec:S-T}.}
  \label{fig:matter vary alpha 2}
\end{subfigure}

\begin{subfigure}{\textwidth}
  \centering
  \includegraphics[scale=0.45]{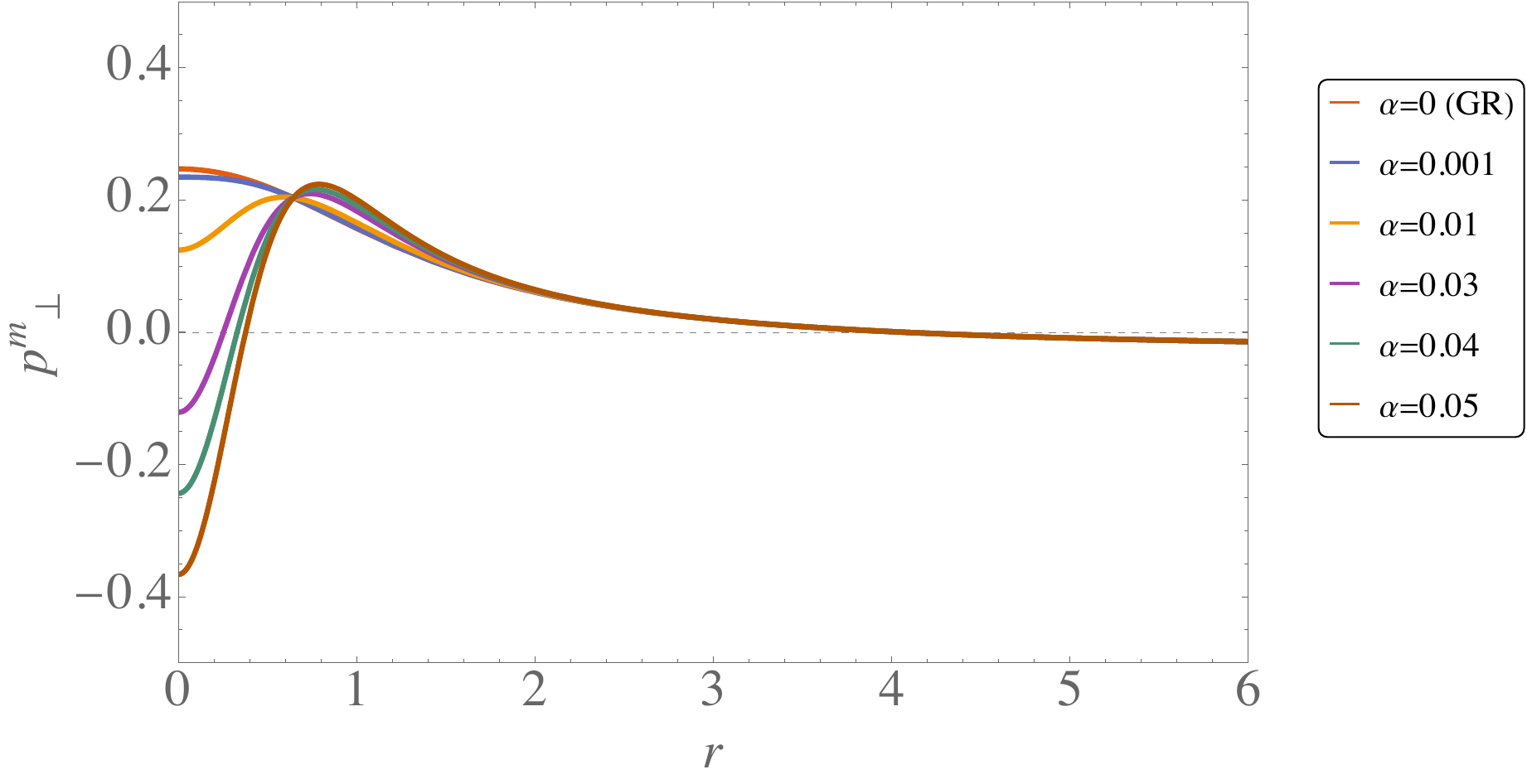}
  \caption{Tangential pressure of the matter fluid for different values of $\alpha$ when considering the (IS-TIV) geometry in Sec. \ref{sec:S-T}.}
  \label{fig:matter vary alpha 3}
\end{subfigure}
\caption{Matter fluid solutions for varied values of $\alpha$ using the same parameter values as in Fig. \ref{fig:starobinsky soln 1}. These are generated for the (IS-TIV) geometry in Sec. \ref{sec:S-T}. Here, $r$ is the normalized area radius, i.e. $r/r_{0}$ with $r_{0}=1$.}
\label{fig: varied-alpha 1}
\end{figure}

\begin{figure}
\begin{subfigure}{\textwidth}
  \centering
  \includegraphics[scale=0.45]{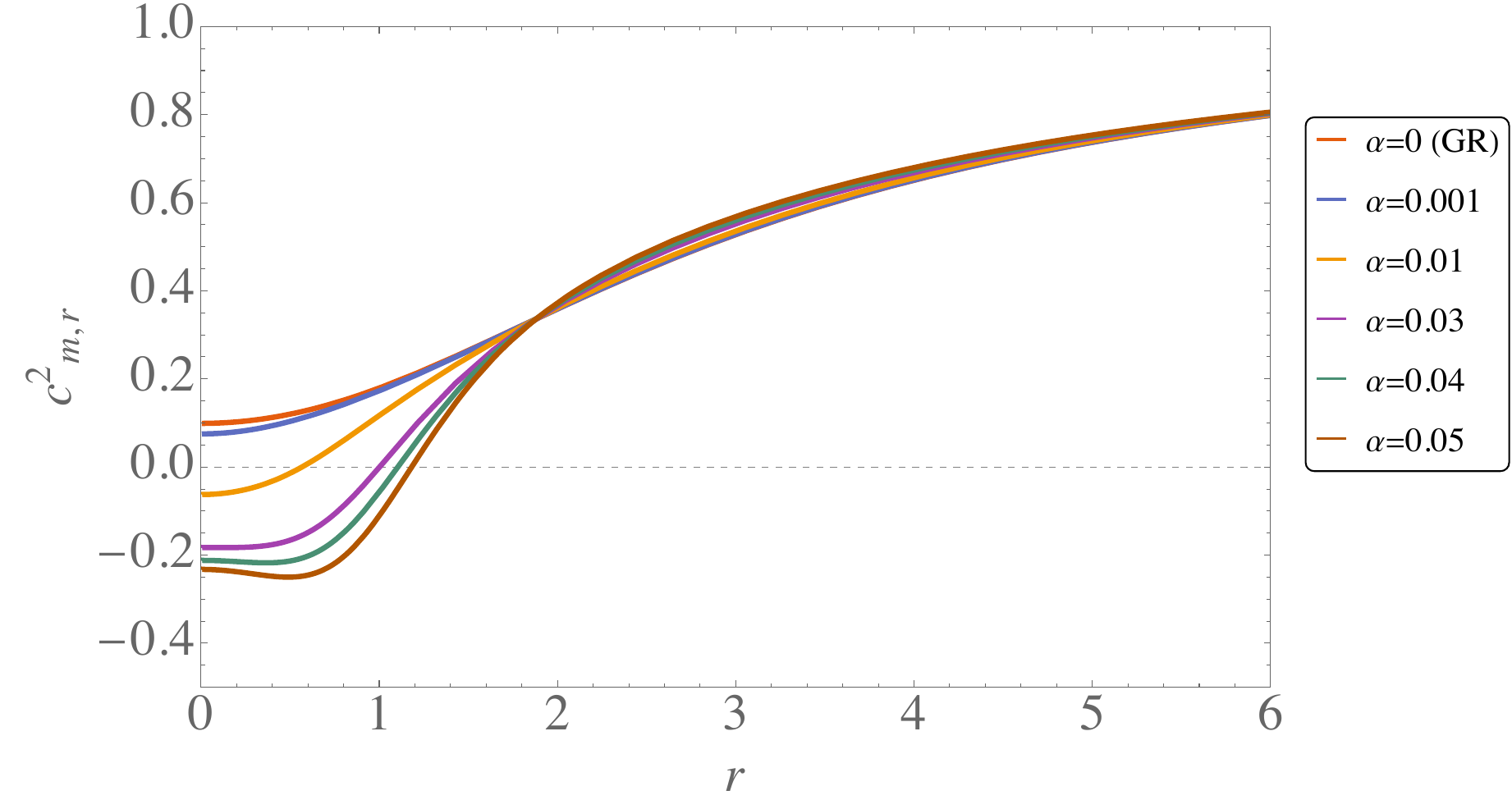}
  \caption{Radial speed of sound of the matter fluid for different values of $\alpha$ when considering the (IS-TIV) geometry in Sec. \ref{sec:S-T}.}
  \label{fig:matter vary alpha 4}
\end{subfigure}

\begin{subfigure}{\textwidth}
  \centering
  \includegraphics[scale=0.45]{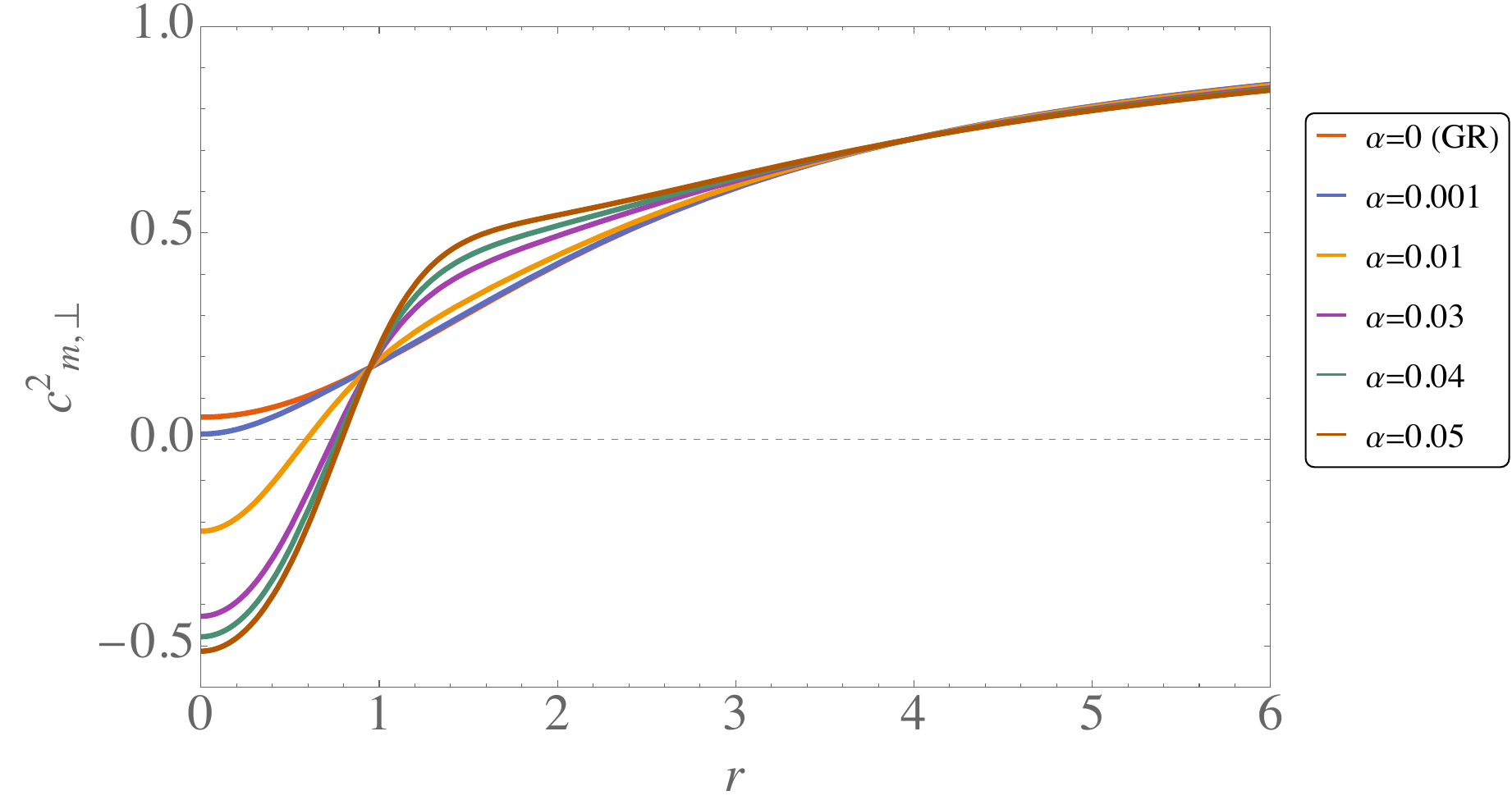}
  \caption{Tangential speed of sound of the matter fluid for different values of $\alpha$ when considering the (IS-TIV) geometry in Sec. \ref{sec:S-T}.}
  \label{fig:matter vary alpha 5}
\end{subfigure}
\caption{The radial and tangential speed of sound of the matter fluid for varied values of $\alpha$ using the same parameter values as in Fig. \ref{fig:starobinsky soln 1}. These are generated for the (IS-TIV) geometry in Sec. \ref{sec:S-T}. Here, $r$ is the normalized area radius, i.e. $r/r_{0}$ with $r_{0}=1$.}
\label{fig: varied-alpha 2}
\end{figure}

\begin{figure}[h]
\centering
\begin{subfigure}{\textwidth}
  \centering
  \includegraphics[scale=0.38]{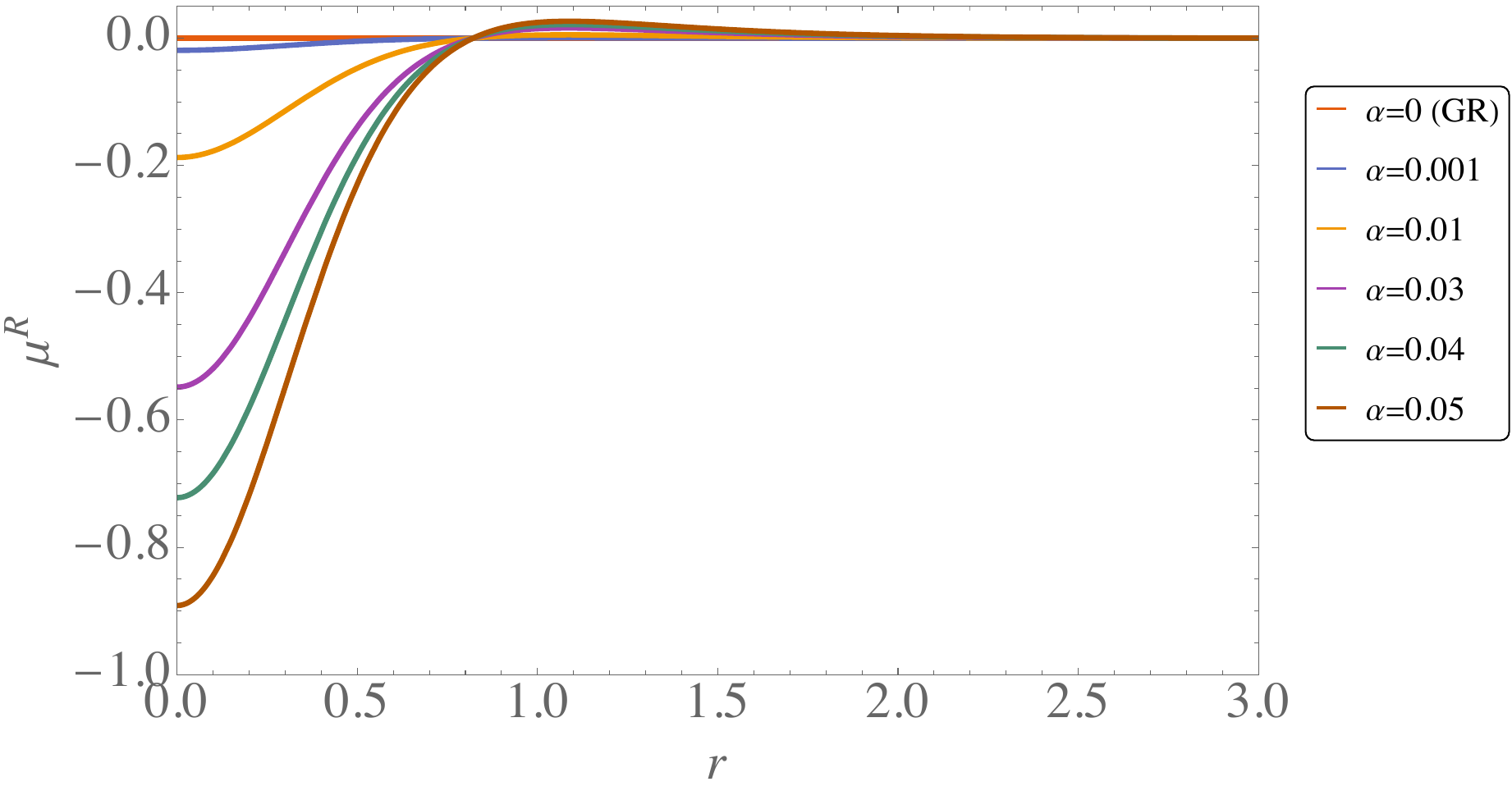}
  \caption{Energy density of the curvature fluid for varied values of $\alpha$.}
  \label{fig:matter vary alpha 6}
\end{subfigure}%

\begin{subfigure}{\textwidth}
  \centering
  \includegraphics[scale=0.38]{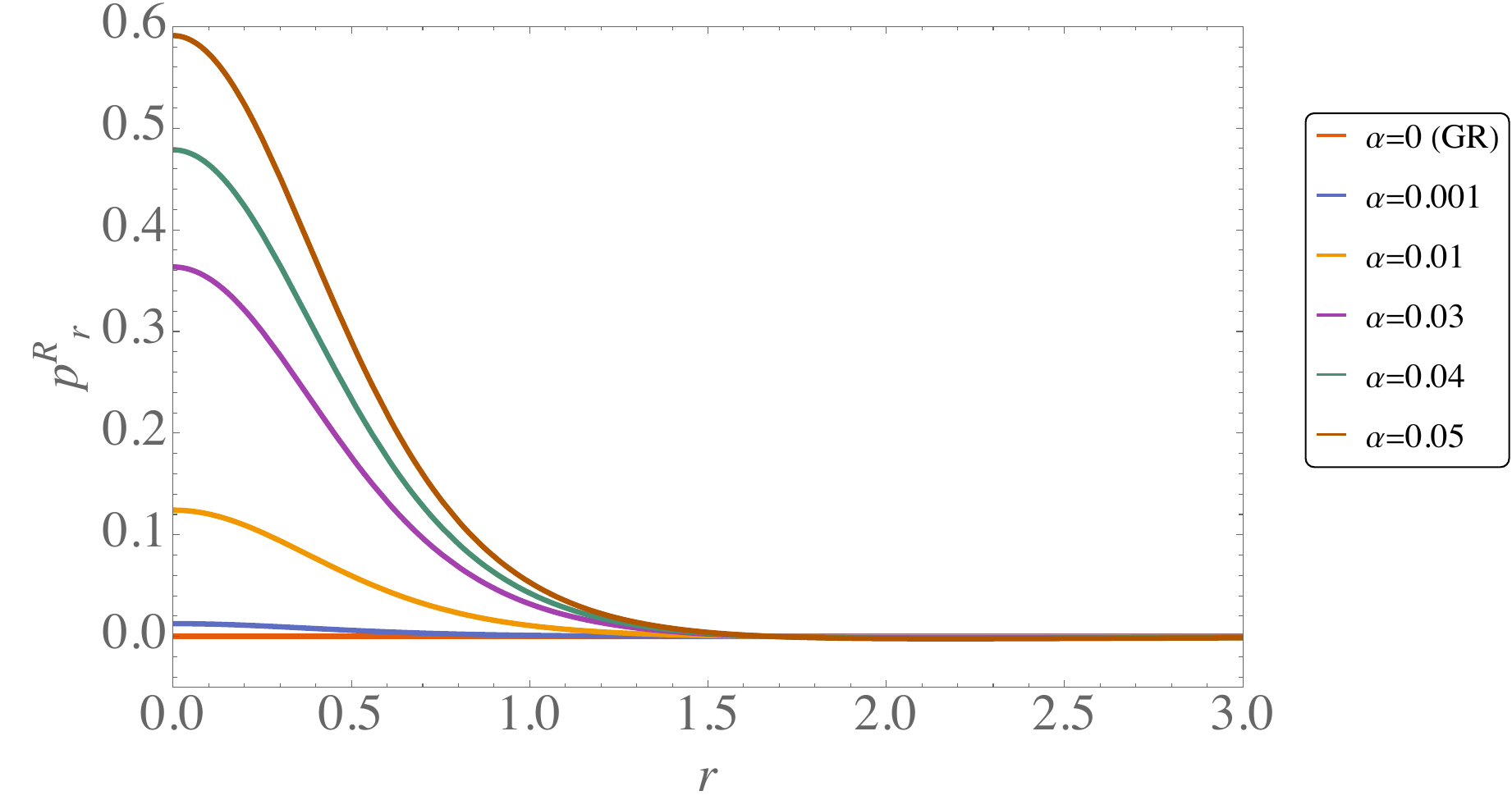}
  \caption{Radial pressure of the curvature fluid for varied values of $\alpha$.}
  \label{fig:matter vary alpha 7}
\end{subfigure}

\begin{subfigure}{\textwidth}
  \centering
  \includegraphics[scale=0.38]{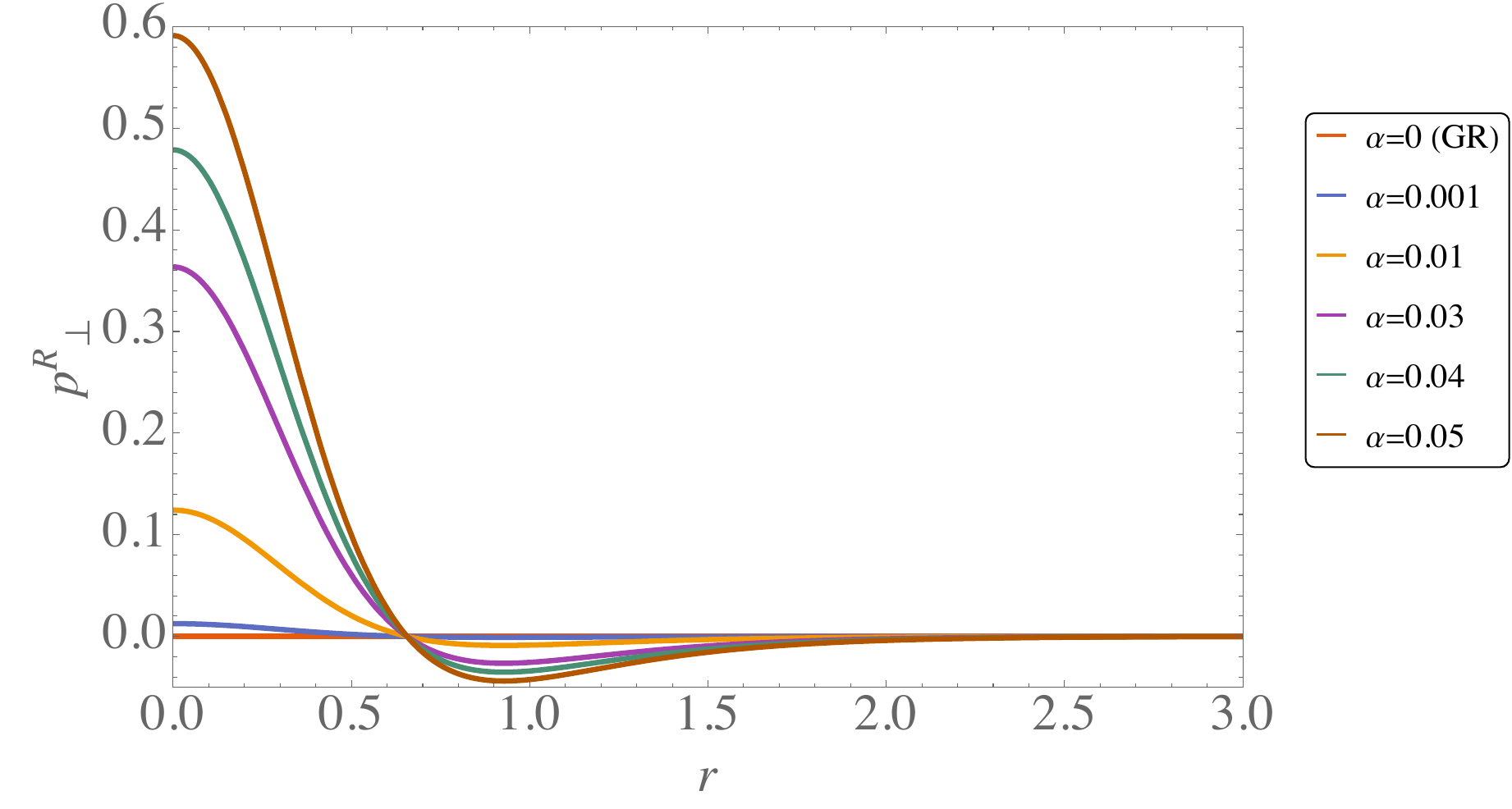}
  \caption{Tangential pressure of the curvature fluid for varied values of $\alpha$.}
  \label{fig:matter vary alpha 8}
\end{subfigure}
\caption{Curvature fluid solutions for varied values of $\alpha$ using the same parameter values as in Fig. \ref{fig:starobinsky soln 1}. These are generated for the (IS-TIV) geometry in Sec. \ref{sec:S-T}. Here, $r$ is the normalized area radius, i.e. $r/r_{0}$ with $r_{0}=1$.}
\label{fig: varied-alpha 3}
\end{figure}

In Fig. \ref{fig: param space test}, we illustrate the likelihood of finding a set of parameters that satisfies the causal condition for the radial matter fluid alone by performing a ``perturbation'' away from the values of the parameters for a solution satisfying the physical conditions. In Figs. \ref{fig:matter vary alpha 1}, \ref{fig: varied-alpha 1}, \ref{fig: varied-alpha 2}, and \ref{fig: varied-alpha 3}, we present the fluid descriptions of the matter and effective curvature fluid for various values of $\alpha$. 
We notice that $\alpha$ affects the slope of the energy density (Fig. \ref{fig:matter vary alpha 1}) and pressures, particularly of the matter fluid, near the center of the stellar object (Fig. \ref{fig: varied-alpha 1}). Therefore, we see that for an increasing value of $\alpha$, $c_{m,r}^{2}$ and $c_{\perp,r}^{2}$, shifts towards being negative around the center of the stellar object. Interestingly, the matter energy density converges to the same value towards the boundary of the stellar object for various values of $\alpha$.


\section{Reconstruction with a generic interior metric}\label{sec: CD-TIV}
In this section, we will apply the reconstruction technique by considering a generic metric anzats.  We chose the component $k_{1}$ as a generalization of the anzats of the corresponding term of the Tolman IV solution obtained by adding a quartic term. We choose the component $k_{2}$  of the metric as a rational function, and we keep it as general as possible. More specifically, we assume a line element of the form \eqref{generic metric} with
\begin{equation}\label{eq: quartic metric}
    k_{1}(r) = 1+\mathfrak{D}_{1}r^{2}+\mathfrak{D}_{2}r^{4}, ~~~~
    k_{2}(r) = \frac{1+\mathfrak{D}_{3}r^{2}}{1+\mathfrak{D}_{4}r^{2}+\mathfrak{D}_{5}r^{4}}.
\end{equation}
The total fluid quantities, in terms of the general metric coefficients, are 
\begin{align}
    \mu^{\text{tot}} &= \frac{rk_{2}'(r)+k_{2}(r){2}-k_{2}(r)}{r^{2}k_{2}(r)^{2}},\\
    p^{\text{tot}}_{r} &= \frac{rk_{1}'(r)-k_{1}(r)k_{2}(r)+k_{1}(r)}{r^{2}k_{1}(r)k_{2}(r)},\\
    p^{\text{tot}}_{\perp} &= \frac{k_{1}''(r)}{2k_{1}(r)k_{2}(r)} - \frac{k_{1}'(r)k_{2}'(r)}{4k_{1}(r)k_{2}(r)^{2}} - \frac{k_{1}'(r)^{2}}{4k_{1}(r)^{2}k_{2}(r)} - \frac{k_{2}'(r)}{2rk_{2}(r)^{2}} + \frac{k_{1}'(r)}{2rk_{1}(r)k_{2}(r)}.
\end{align}
The complete expressions in terms of the parameters is in the Appendix \ref{appendix: solution 2 case II}. 

In order to find realistic solutions, the junction conditions are implemented by setting $R(r_{b})=\hat{R}(r_{b})=p^{m}_{r}(r_{b})=0$, where $r_{b}$ is where we set the boundary of the star. This allows us to eliminate and constrain parameter dependencies.

By implementing the junction conditions for a smooth matching, the number of parameter dependencies is reduced to only three : $\mathfrak{D}_{1}$, $\mathfrak{D}_{2}$, and $\alpha$. We follow the same procedure to finding solutions to the TOV equations, outlined in Sec. \ref{sec:S-T}. The full expressions for the thermodynamical quantities in terms of the metric coefficients are given in \ref{appendix: solution 2 case II}. 

\begin{figure}
    \centering
    \includegraphics[scale=0.55]{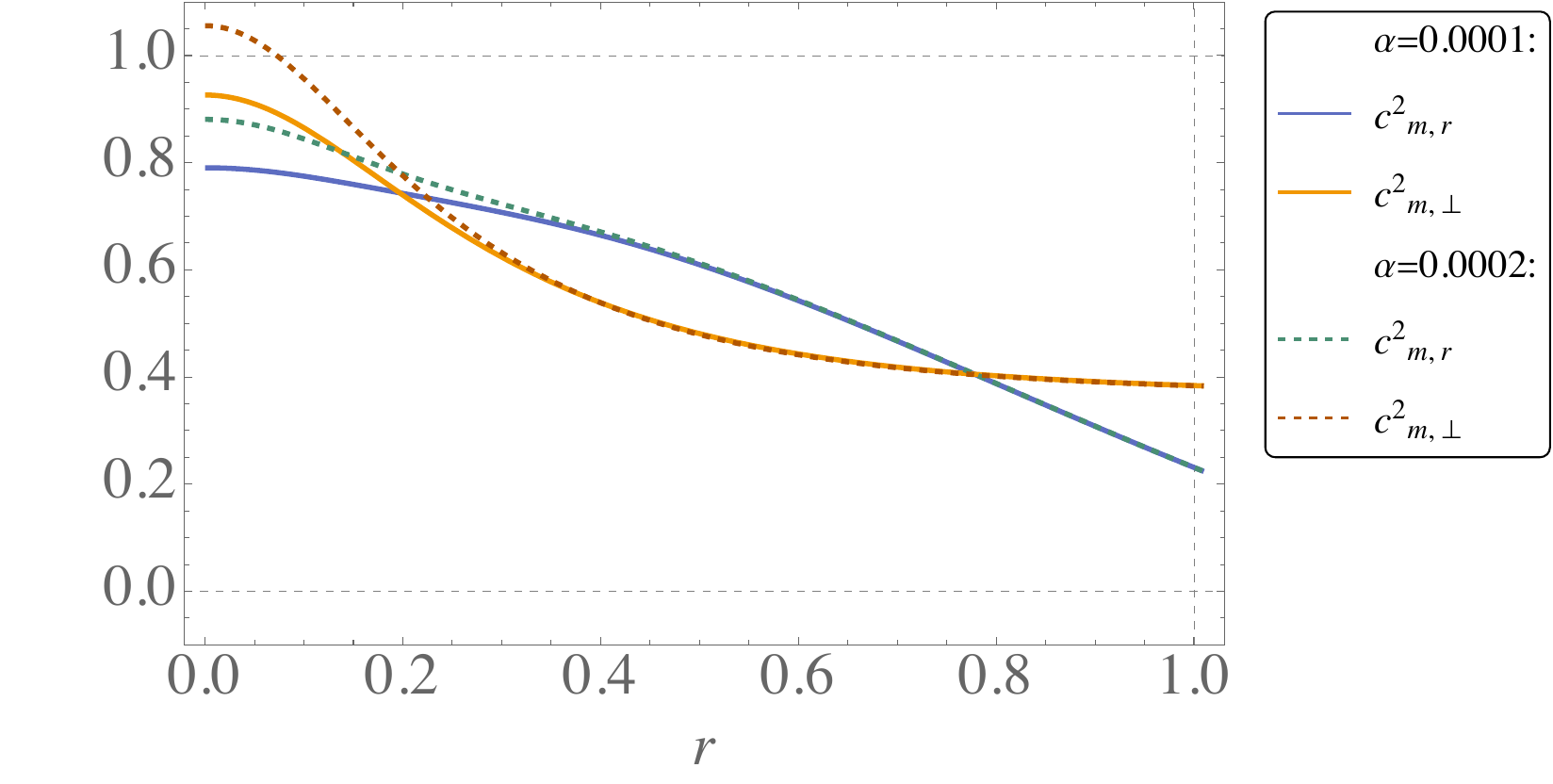}
    \caption{The radial and orthogonal speeds of sound of baryonic matter for the solution with a generic interior metric in Sec. \ref{sec: CD-TIV}. Then this solution represents a stellar object that is highly compact at the core ($r=0$) since $c^{2}_{m,r}\simeq0.8$. We include a small incremental change to $\alpha$ to illustrate how sensitive the speed of sound is to a change in the $\alpha$ parameter, as we have seen in Sec. \ref{sec:S-T}. For $\alpha>0.00015$, we find $c^{2}_{m,\perp}>1$ at the center of the stellar object.}
    \label{fig: caseII matter sos}
\end{figure}

Figure \ref{fig:quartic A met coeff 2} and \ref{fig: caseII matter sos} show the radial behavior of the baryonic matter for particular values of the parameters. This case admits a solution that satisfies the physical conditions in Sec. \ref{sec:fluid constraints} and that a smooth matching to the surfaces is possible for quadratic $f(R)$ models with a positive value of its model parameter, $\alpha=0.0001$ (i.e., $\alpha\sim10^{6}~cm^{2}$, which is compatible with the constraint found in \cite{ConstraintsonfRviaNeutronStars}).

\begin{figure}
    \centering
    \includegraphics[scale=0.55]{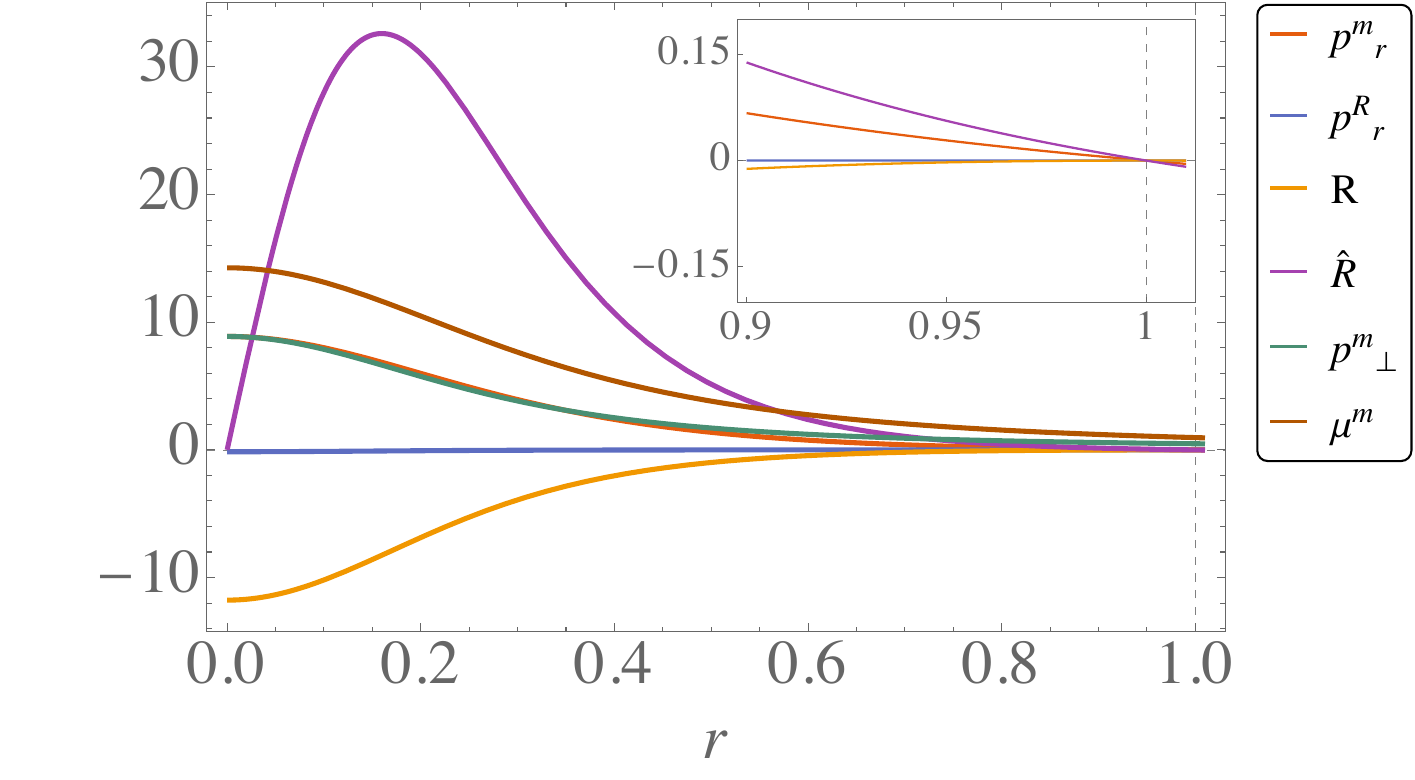}
    \caption{Fluid solutions to baryonic matter for a quartic $f(R)$ model with $\alpha=0.0001$, $\mathfrak{D}_{1}=6.8$ and $\mathfrak{D}_{2}=10$. The boundary of the star is at $r=r_{b}=1$, where $r$ is the normalized area radius (i.e. $r/r_{0}$ with $r_{0}=1$). This solution shows a smooth matching as outlined in Sec. \ref{sec:junction conditions}, and corresponds to the generic interior metric case considered in Sec. \ref{sec: CD-TIV}.}
    \label{fig:quartic A met coeff 2}
\end{figure}
\begin{figure}
  \centering
  \includegraphics[scale=0.55]{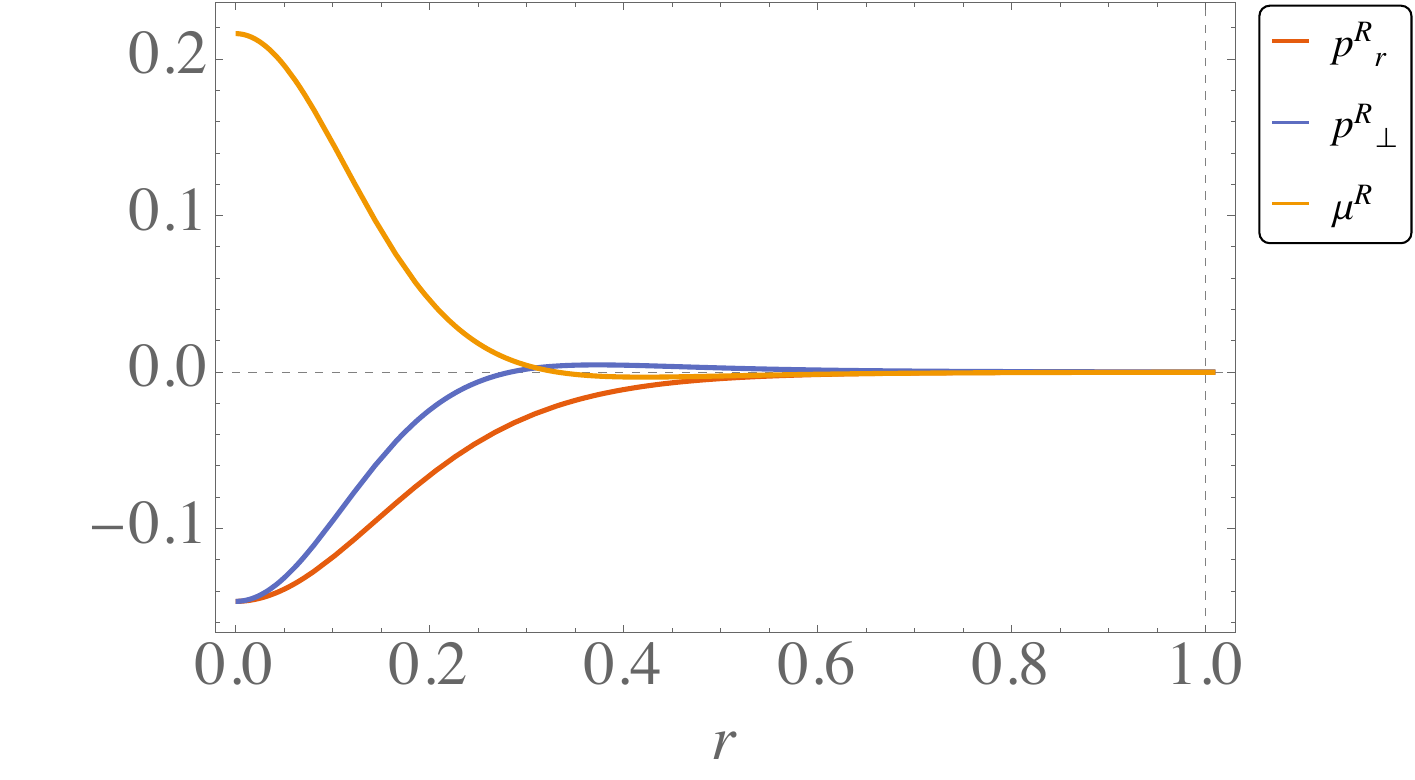}
\caption{Solutions to the curvature fluid for a quartic metric, $f(R)$ model with $\alpha=0.0001$, $\mathfrak{D}_{1}=6.8$, and $\mathfrak{D}_{2}=10$. The boundary of the star is at $r=r_{b}=1$, where $r$ is the normalized area radius (i.e. $r/r_{0}$ with $r_{0}=1$). This solution corresponds to the solution with a generic interior metric in Sec. \ref{sec: CD-TIV}.}
\label{fig: generic metric case 2 curvature}
\end{figure}

\begin{figure}
  \centering
  \includegraphics[scale=0.82]{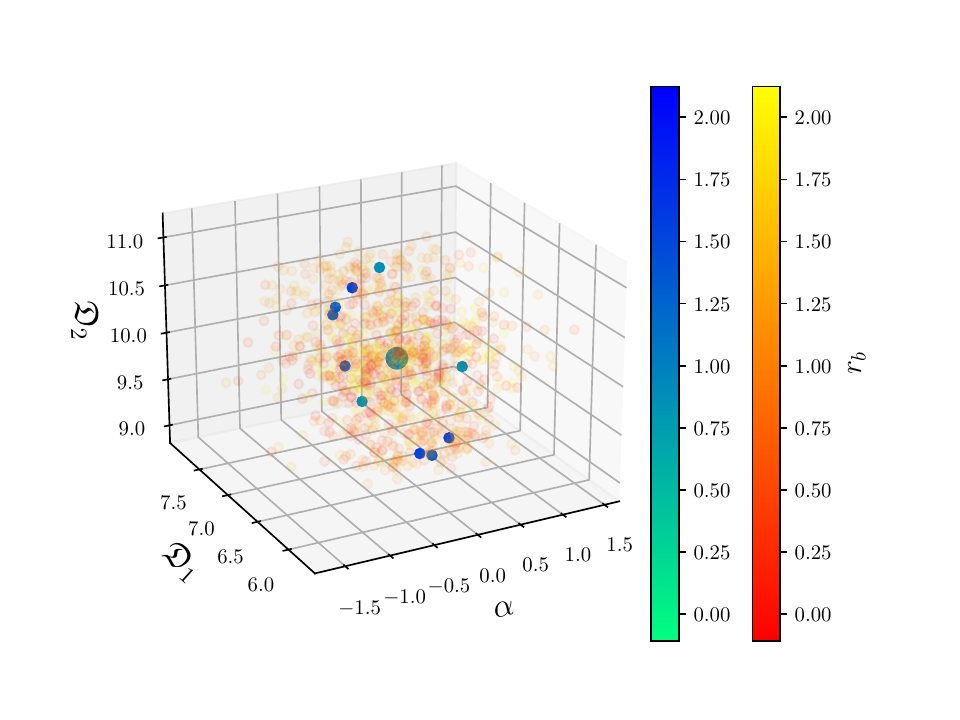}
  \caption{Parameter space plot for the radial, squared speed of sound of baryonic matter for the generic interior metric in Sec. \ref{sec: CD-TIV}. The faint points are the random parameter values generated and the darker, shades of green points are the ones satisfying $0\leq c^{2}_{m,r}\leq1$. Compared to the case of Fig. \ref{fig: param space test}, the number of random sets of parameter values are doubled and only 1\% of them satisfy the causal condition $c^{2}_{m,r}$.}
  \label{fig: param space test 2}
\end{figure}

\begin{figure}[h!]
  \centering
  \includegraphics[scale=0.35]{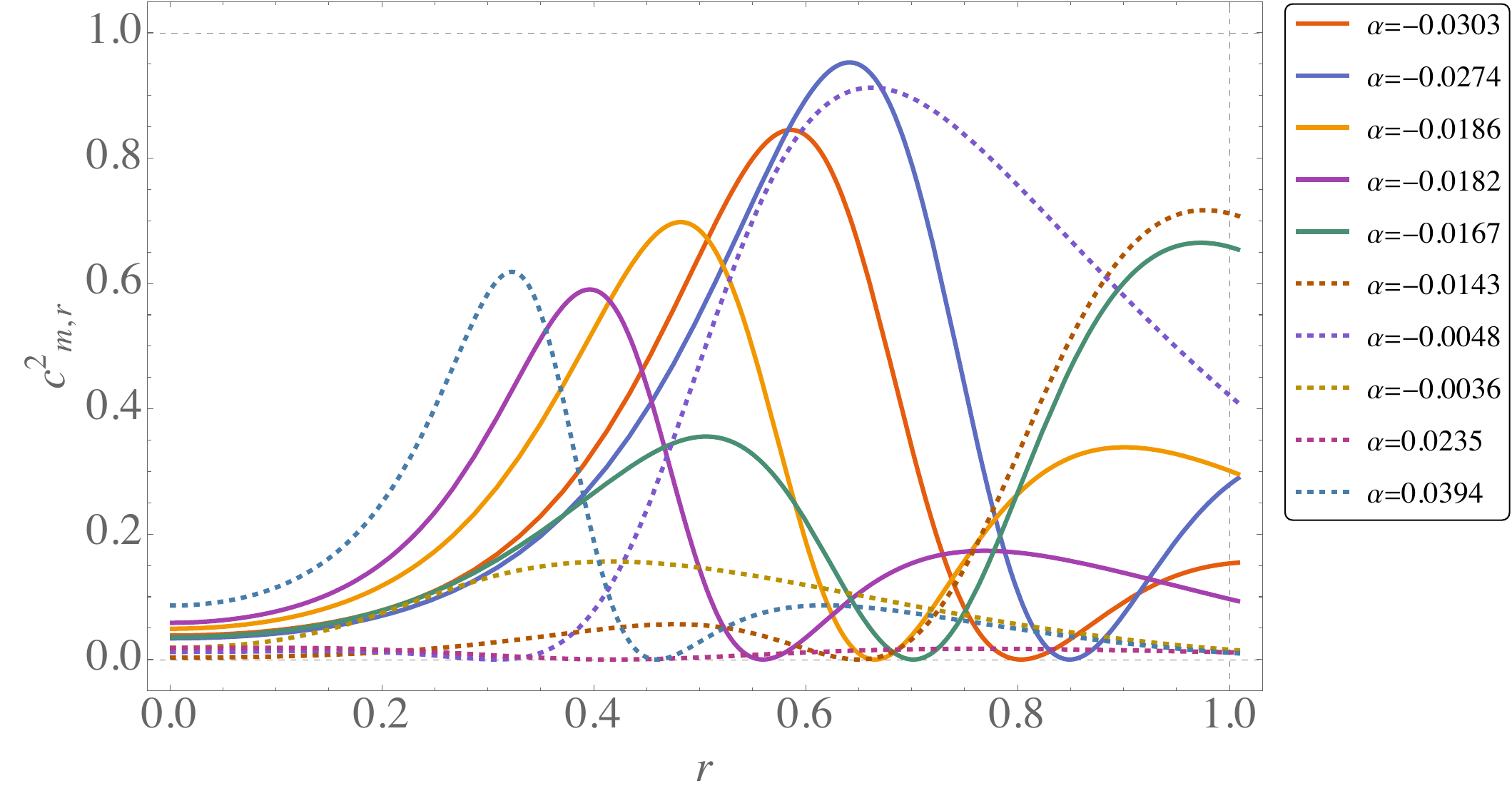}
  \caption{This shows the radial speed of sound squared of baryonic matter, using the set of parameter values in Fig. \ref{fig: param space test 2} that satisfies the causal condition. The boundary of the star is at $r=r_{b}=1$, where $r$ is the normalized area radius (i.e. $r/r_{0}$ with $r_{0}=1$).}
  \label{fig: sos paramspace 2}
\end{figure}

\clearpage
\section{Discussion and Conclusion} \label{sec:conclusion}
We presented a study on the extension of the TOV equations to the case of theories of gravity of order four, which are characterized by a non-linear action in the Ricci scalar: the so-called $f(R)$ theories of gravity.

By employing the (1+1+2) formalism, we have rewritten the TOV equations in a covariant and dimensionless form, valid for any function $f$. This result was achieved by recognizing that $f(R)$ gravity can always be recast as GR plus two effective, non-interacting fluids, one of which is not a perfect fluid.

The generalized TOV equations can then be used as a framework for finding exact, analytical solutions to static spherically symmetric spacetimes, which can describe relativistic stars for $f(R)$ gravity. In this context, the work developed in \cite{NCD,NCD2} can be applied to the search for new exact solutions of the TOV equations. In particular, we have used the so-called reconstruction algorithm of \cite{NCD,NCD2}, in which a solution to the matter fluid can be found by making an ansatz of the description of the metric tensor. As noted in the previous section, not all solutions found this way are physical since the fluid sources must satisfy specific physical requirements such as energy conditions, causality, etc.

Another important issue in constructing meaningful interior solutions is the solution's boundary connection with an exterior spacetime. It is well known that in $f(R)$ gravity Israel's junction conditions are modified, and additional constraints are required to join two spacetimes. These constraints are connected to the degree of freedom carried by the curvature scalar and this makes the search for exact solutions to relativistic stars in $f(R)$ gravity much harder than in GR. In addition, when there is a mismatch in the geometry, quadratic models, like the ones we considered, can present a double layer. This structure, whose physics is not yet well understood, is similar to the dipole layer that forms at the interface between two charged fluids. Our exact solutions allow us to characterize double layers exactly and, as such, can be used to improve the understanding of their physics.

As a first example, in Sec. \ref{sec:S-T}, we choose $k_{1}$ to be described by the $(0,0)$ interior Schwarzschild metric coefficient, and $k_{2}$ to be represented by the $(1,1)$ Tolman IV metric coefficient (the same reasoning is applied in Sec. \ref{sec: CD-TIV}). This choice was motivated by the attempt to preserve the physically relevant features of these metrics in terms of the Newtonian limit and simplicity. In addition, as shown in \cite{NCD,NCD2}, this hybrid metric describes an anisotropic fluid, thereby offering an ideal framework for $f(R)$ gravity. 

 Our analysis shows that there are sets of parameters for which the metric Eq. \eqref{generic metric} with coefficients Eq. \eqref{eq: metric coeffs 1 const density tolman 4} satisfies the physical conditions of Sec. \ref{sec:fluid constraints} and therefore corresponds to a physical relativistic star. However, this solution does not match smoothly with a Schwarzschild exterior, and a shell with double layer has to be introduced to regularize the spacetime. We have determined that this is the case for the solution we have found, and therefore we had the chance to explore in detail the working of the double layer. Analysis of the properties of the tensor $S_{ab}$ shows that the solution requires a shell with a tension to be stable. The different correction introduced by the double layer change the situation is a small but intricate way. In fact, the tensor $\bar{S}_{ab}$, contains a positive radial pressure components and a  positive correction to the orthogonal pressure, which tend to reduce the tension necessary for stability. On the other hand the double layer contributes with a energy density and orthogonal pressure which are proportional to each other and to the parameter $\alpha$. Notice that the sign of $\alpha$ regulates the double layer contribution, so that for a double layer that has standard fluid properties $\alpha$ must be positive. This is the same condition which is known to guarantee that the mass of the scalaron, the additional gravitational scalar degree of freedom of Starobinsky's model, is positive and leaves us to wonder if these two aspects of the physics of this theory are indeed connected.

We found it worth exploring the behavior of the solutions as the parameters change. For example, in the case of a coupling constant $\alpha\ll1$ with respect to the gravitational action, comparing the behavior of the tangential and radial pressures, we see that with these parameters, this solution represents a ``quasi-isotropic object'', similar to the ones found in \cite{sante-TOV-aniso-GR1} for the single fluid case(see Fig.\ref{fig: ricci-scalar-tot-pressure}). This type of object occurs when the radial and tangential pressures behave similarly.  Still, anisotropies influence the behavior of other physical parameters. For example, in our solution, the radial and orthogonal sound speeds differ in behavior. The tangential speed of sound, in particular, has a minimum around the center of the stellar object, corresponding to a maximum of the anisotropy. Note that the effective fluid generated by the curvature invariants appears to have an energy density and pressure considerably smaller than the ones of baryonic matter. Therefore, the new solution represents an object mostly made of baryonic matter whose structure would differ from a corresponding GR object. The curvature fluid also presents a positive energy density and pressure (see Fig. \ref{fig:starobinsky soln 3}).

 As the value of $\alpha$ increases, we see that the pressure of the curvature fluid increases together with the energy density of baryonic matter. Still, the pressures of the matter fluid generally decreases (see Fig. \ref{fig:matter vary alpha 1} and \ref{fig: varied-alpha 1}). Interestingly, the speeds of sound of matter change dramatically close to the center, becoming quickly negative (see Fig. \ref{fig: varied-alpha 2}). 

In Sec. \ref{sec: CD-TIV}, we performed the reconstruction starting from a completely general metric expressed in terms of polynomial and rational functions. These functions have the advantage of offering a sufficient number of parameters and also reducing the growth in complexity of the TOV equations. For the (0,0) component of the metric, we considered a quadratic polynomial that is related by the constant density-Tolman IV metric solution with a fourth-order correction to the (0,0) component of the metric. 

With this extension, we found a solution with a smooth matching of the boundary surface for $\alpha=0.0001$, and its baryonic matter profile seems fairly standard with respect to other known solutions of the TOV equations. Its pressures, energy density and speeds of sound profile have a monotonically decreasing profile. However, compared to the solution in Sec. \ref{sec:S-T}, Fig. \ref{fig: caseII matter sos} shows a stellar object that has more of a compact core, i.e. $c^{2}_{m,r}(r=0)\simeq0.8$.

It is interesting to compare the relative magnitude between the matter quantities and the effective curvature fluid quantities. The effective curvature quantities in Sec. \ref{sec: CD-TIV}, Fig. \ref{fig: generic metric case 2 curvature}, are less but comparable to the baryonic matter solutions as opposed to the solution in Sec. \ref{sec:S-T} where the effective curvature quantities are considerably smaller. This feature reminds us that a small fourth-order perturbation in the action does not necessarily translate into a small deviation from the properties of analogous gravitational systems in GR.

Figure \ref{fig: param space test 2} shows that a small deviation from the solution results in $1\%$ of the parameter value sets satisfying the causal condition of $c^{2}_{m,r}$. Of these $1\%$ parameter value sets, we notice, in  Fig. \ref{fig: sos paramspace 2}, oscillatory behavior in the speeds of sound. The presence of these oscillations could be an indicator of the existence of a potential instability of this solution, which could only be confirmed through a detailed analysis of the perturbations.
On the other hand,  the tangential speed of sound is not always well-defined. For example, in Fig. \ref{fig: caseII matter sos}, we see that there are value of the parameters in which this quantity would violate causality.

The results above suggest that the structure of a relativistic star in $f(R)$ gravity and its response to parameter variations can be very different from those in the GR case. These results suggest that modifying GR might cause important physical structures and composition differences that might one day become measurable. 

Overall, our work shows that analytical approaches can describe astrophysical phenomena in $f(R)$ gravity and that these solutions possess the correct physical features for a quadratic model of $f(R)$ Future work will be dedicated to improving our understanding of the properties of these solutions, with particular emphasis on the observable features that might constitute a signature to test higher-order corrections to the gravitational action of relativistic stars. Specifically, the mass-radius relation of our solution and its maximum mass limit deserves particular attention. Investigating other functional forms of $f(R)$ would also be worthwhile, in particular the Hu-Sawicki model \cite{hu-sawicki} and the $R^{n}$ model where perturbative effects on $R$ can be studied by considering $n=1+\delta$ \cite{CliftonThesis}.

\section{Acknowledgments} 
This work was supported in part by the Italian Ministry of
Foreign Affairs and International Cooperation, grant number M00023. The work of S. C. has been carried out in the framework of activities of the INFN Research Project QGSKY. M. C. acknowledges financial support from the National Research Foundation of South Africa, a holder of the Scarce-Skills Ph.D. Scholarship, and the University of Cape Town Science Faculty Ph.D. Scholarship during which this research was performed. N. F. N. acknowledges funding from the Oppenheimer Memorial Trust. P. K. S. D. acknowledges funding from First Rand Bank (SA) and the National Research Foundation of South Africa for an Italy-South Africa bilateral grant ITAL2204214155. 

\section{Data availability}
No new data were created or analysed in this study.

\appendix

\section{Interior Schwarzschild-Tolman IV spacetime}\label{appendix2}
Here, we give the full expressions for the solution given in Sec. \ref{sec:S-T} and the quantities expressed on the surface $\mathcal{S}$ describing the shell and the double layer. 
\subsection{Full expressions for the quantities evaluated at the junction}\label{eq: full boundary quantities}
\begin{equation} [\phi]^{+}_{-}\equiv\phi^{\mathcal{S}} = \frac{\sqrt{A^2+r_b^2} \sqrt{\mathcal{R}^2-r_b^2}}{\mathcal{R} \sqrt{A^2+2 r_b^2}},
\end{equation}

\begin{equation}
    [\mathcal{A}]^{+}_{-} = \mathcal{A}^{\mathcal{S}} = \frac{\mu _1 r_b^2 \sqrt{A^2+r_b^2} \sqrt{\mathcal{R}^2-r_b^2}}{\mathcal{R} \sqrt{A^2+2 r_b^2} \left(-2 c_1 \sqrt{3-\mu _1 r_b^2}+2 \mu _1 r_b^2-6\right)},
\end{equation}

\begin{equation}
    [X]^{+}_{-} = X^{\mathcal{S}} = -\frac{2 r_b \sqrt{\mathcal{R}^2-r_b^2} \sqrt{A^2+r_b^2} \left(-2 x_{11} +2 x_{12} -c_1^2 \sqrt{3-r_b^2 \mu _1} x_{13} + c_1 x_{10}\right)}{\mathcal{R}^3 \left(A^2+2 r_b^2\right){}^{7/2} \left(3-r_b^2 \mu _1\right){}^{3/2} \left(\mu _1 r_b^2-c_1 \sqrt{3-r_b^2 \mu _1}-3\right){}^3},
\end{equation}
where
\begin{align}
    x_{1} &= 2 \mathcal{R}^2 A^6+12 \mathcal{R}^2 r_b^2 A^4+\left(4 \mathcal{R}^2-6 A^2\right) r_b^6+\left(6 A^2 \mathcal{R}^2-9 A^4\right) r_b^4,\\
    x_{2} &= 8 r_b^8-2 \left(A^2-2 \mathcal{R}^2\right) r_b^6+\left(6 A^2 \mathcal{R}^2-15 A^4\right) r_b^4+2 \left(A^6+13 \mathcal{R}^2 A^4\right) r_b^2+4 A^6 \mathcal{R}^2,\\
    x_{3} &= 16 r_b^6+6 \left(3 A^2-2 \mathcal{R}^2\right) r_b^4-\left(A^4+26 \mathcal{R}^2 A^2\right) r_b^2+4 A^4 \left(A^2+\mathcal{R}^2\right),\\
    x_{4} &= 32 r_b^8+2 \left(A^2-26 \mathcal{R}^2\right) r_b^6-\left(71 A^4+130 \mathcal{R}^2 A^2\right) r_b^4+\left(8 A^6+38 \mathcal{R}^2 A^4\right) r_b^2+5 A^6 \mathcal{R}^2,\\
    x_{5} &= 16 r_b^6+6 \left(A^2-6 \mathcal{R}^2\right) r_b^4-\left(31 A^4+86 \mathcal{R}^2 A^2\right) r_b^2+4 A^4 \left(A^2+\mathcal{R}^2\right),\\
    x_{6} &= 12 r_b^8-6 A^2 r_b^6-2 \left(16 A^4+5 \mathcal{R}^2 A^2\right) r_b^4+3 \left(A^6+13 \mathcal{R}^2 A^4\right) r_b^2+6 A^6 \mathcal{R}^2,\\
    x_{7} &= 60 r_b^8+4 \left(A^2-5 \mathcal{R}^2\right) r_b^6-\left(103 A^4+68 \mathcal{R}^2 A^2\right) r_b^4+3 \left(5 A^6+43 \mathcal{R}^2 A^4\right) r_b^2+19 A^6 \mathcal{R}^2,\\
    x_{8} &= 112 r_b^8+\left(70 A^2-92 \mathcal{R}^2\right) r_b^6-3 \left(35 A^4+74 \mathcal{R}^2 A^2\right) r_b^4\nonumber\\
    &+2 \left(14 A^6+53 \mathcal{R}^2 A^4\right) r_b^2+13 A^6 \mathcal{R}^2,\\
    x_{9} &= 4 r_b^6+3 \left(A^2-2 \mathcal{R}^2\right) r_b^4-2 \left(2 A^4+7 \mathcal{R}^2 A^2\right) r_b^2+A^4 \left(A^2+\mathcal{R}^2\right),\\
    x_{10} &= r_b^4 x_{6} \mu _1^4-3 r_b^2 x_{7} \mu _1^3+9 x_{8} \mu _1^2-432 x_{9} \mu _1-486 \left(A^2+2 \mathcal{R}^2\right) \left(5 A^2+2 r_b^2\right),\\
    x_{11} &= \left(r_b^2 x_{1} \mu _1^3-3 x_{2} \mu _1^2+9 x_{3} \mu _1+27 \left(A^2+2 \mathcal{R}^2\right) \left(5 A^2+2 r_b^2\right)\right)\left(3-r_b^2 \mu _1\right){}^{3/2},\\
    x_{12} &= \left(A^2+2 \mathcal{R}^2\right) c_1^3 \left(5 A^2+2 r_b^2\right) \left(r_b^2 \mu _1-3\right){}^3,\\
    x_{13} &= r_b^2 \left(12 r_b^8+2 \left(A^2-8 \mathcal{R}^2\right) r_b^6-6 \left(4 A^4+7 \mathcal{R}^2 A^2\right) r_b^4 +3 \left(A^6+5 \mathcal{R}^2 A^4\right) r_b^2+2 A^6 \mathcal{R}^2\right) \mu _1^3\nonumber\\ 
    &- 3 x_{4} \mu _1^2+18 x_{5} \mu _1+162 \left(A^2+2 \mathcal{R}^2\right) \left(5 A^2+2 r_b^2\right).
\end{align}

The expression for the Ricci scalar along the surface $\mathcal{S}$ is Eq. \eqref{eq: final Ricci} evaluated at $r_{b}$ and the parameter values used in the Figs. \ref{fig:starobinsky soln 1}--\ref{fig: param space test}.

\subsection{Ricci scalar:}
\begin{align}\label{eq: final Ricci}
    R(r) = \frac{1}{B}&\left(2 \left(A^4 \left(3 c_1^2 \mathit{z}{}^{3}+c_1 a_{1}+\mathit{z} a_{2}\right)+A^2 \left(c_1^2 \mathit{z}{}^{3} \left(7 r^2+3 \mathcal{R}^2\right)+c_1 a_{3} + \mathit{z} a_{4}\right)\right.\right.\nonumber\\ 
    &\left.\left.+ 2 r^2 \left(c_1^2 \mathit{z}{}^{3} \left(3 r^2+\mathcal{R}^2\right)+3 c_1 a_{5} +\mathit{z} a_{6}\right)\right)\right),
\end{align}
where $\mathit{z}=\sqrt{3-\mu _1 r^2}$, and
\begin{align}
    &a_{1} = 9 \mu _1^2 r^4-2 \mu _1 r^2 \left(\mu _1 \mathcal{R}^2+24\right)+9 \left(\mu _1 \mathcal{R}^2+6\right),\\
    &a_{2} = 6 \mu _1^2 r^4-2 \mu _1 r^2 \left(\mu _1 \mathcal{R}^2+15\right)+9 \left(\mu _1 \mathcal{R}^2+3\right),\\
    &a_{3} = 22 \mu _1^2 r^6+\mu _1 r^4 \left(\mu _1 \mathcal{R}^2-117\right)-6 r^2 \left(2 \mu _1 \mathcal{R}^2-21\right)+54 \mathcal{R}^2,\\
    &a_{4} = 15 \mu _1^2 r^6-\mu _1 r^4 \left(2 \mu _1 \mathcal{R}^2+75\right)+r^2 \left(6 \mu _1 \mathcal{R}^2+63\right)+27 \mathcal{R}^2,\\
    &a_{5} = 3 \mu _1^2 r^6-16 \mu _1 r^4+r^2 \left(18-\mu _1 \mathcal{R}^2\right)+6 \mathcal{R}^2,\\
    &a_{6} = 6 \mu _1^2 r^6-\mu _1 r^4 \left(\mu _1 \mathcal{R}^2+30\right)+3 r^2 \left(\mu _1 \mathcal{R}^2+9\right)+9 \mathcal{R}^2,\\
    &B= \mathcal{R}^2 \left(A^2+2 r^2\right)^2 \mathit{z} \left(c_1 \mathit{z}-\mu _1 r^2+3\right){}^2.
\end{align}

\subsection{Total energy density:}
\begin{equation}\label{eq: final tot mu}
    \mu^{\text{tot}}=\frac{3 A^4+A^2 \left(7 r^2+3 \mathcal{R}^2\right)+2 r^2 \left(3 r^2+\mathcal{R}^2\right)}{\mathcal{R}^2 \left(A^2+2 r^2\right)^2}.
\end{equation}

\subsection{Total radial pressure:}
\begin{equation}\label{eq: final tot radial pressure}
    p_{r}^{\text{tot}} = \frac{4 \left(A^2+r^2\right) \left(\mathcal{R}^2-r^2\right)}{r^2 \mathcal{R}^2 \left(A^2+2 r^2\right)} \left(-\frac{\mathcal{R}^2 \left(A^2+2 r^2\right)}{4 \left(A^2+r^2\right) \left(\mathcal{R}^2-r^2\right)}+\frac{\mu _1 r^2}{-2 c_1 \mathit{z}+2 \mu _1 r^2-6}+\frac{1}{4}\right).
\end{equation}

\subsection{Total isotropic pressure:}
\begin{align}\label{eq: final tot iso pressure}
    p^{\text{tot}} = \frac{1}{3B}&\left(-\left(A^4 \left(3 c_1^2 \mathit{z}{}^{3}+2 c_1 b_{1} + \mathit{z} b_{2}\right)\right)-A^2 \left(c_1^2 \mathit{z}{}^{3} \left(7 r^2+3 \mathcal{R}^2\right)+2 c_1 b_{3} + \mathit{z} b_{4}\right)\right.\nonumber\\ 
    &\left.- 2 r^2 \left(c_1^2 \mathit{z}{}^{3} \left(3 r^2+\mathcal{R}^2\right)+2 c_1 b_{5} + 3 \mathit{z} b_{6}\right)\right),
\end{align}
where
\begin{align}
    &b_{1} = \mu _1 \left(6 \mu _1 r^4-2 r^2 \left(\mu _1 \mathcal{R}^2+15\right)+9 \mathcal{R}^2\right)+27,\\
    &b_{2} = \mu _1 r^2 \left(9 \mu _1 r^2-4 \mu _1 \mathcal{R}^2-42\right)+18 \mu _1 \mathcal{R}^2+27,\\
    &b_{3} = r^2 \left(\mu _1 r^2 \left(15 \mu _1 r^2-2 \mu _1 \mathcal{R}^2-75\right)+6 \mu _1 \mathcal{R}^2+63\right)+27 \mathcal{R}^2,\\
    &b_{4} = r^2 \left(\mu _1 r^2 \left(23 \mu _1 r^2-7 \mu _1 \mathcal{R}^2-108\right)+30 \mu _1 \mathcal{R}^2+63\right)+27 \mathcal{R}^2,\\
    &b_{5} = r^2 \left(\mu _1 r^2 \left(6 \mu _1 r^2-\mu _1 \mathcal{R}^2-30\right)+3 \mu _1 \mathcal{R}^2+27\right)+9 \mathcal{R}^2,\\
    &b_{6} = r^2 \left(\mu _1 r^2 \left(3 \mu _1 r^2-\mu _1 \mathcal{R}^2-14\right)+4 \mu _1 \mathcal{R}^2+9\right)+3 \mathcal{R}^2.
\end{align}

\subsection{Total orthogonal pressure:}
\begin{align}\label{eq: final perp pressure}
    p_{\perp}^{\text{tot}} = &\frac{1}{B}\left(A^4 \left(-c_1^2 \mathit{z}{}^{3}+c_1 d_{1} + \mathit{z} d_{2}\right)-A^2 \left(c_1^2 \mathit{z}{}^{3} \left(2 r^2+\mathcal{R}^2\right)+3 c_1 d_{3} + \mathit{z} d_{4}\right)\right.\nonumber\\
    &\left.-2 r^4 \left(c_1^2 \mathit{z}{}^{3} - c_1 d_{1} + \mathit{z} d_{5}\right)\right),
\end{align}
where
\begin{align}
    &d_{1} = \mu _1 r^2 \left(-4 \mu _1 r^2+\mu _1 \mathcal{R}^2+21\right)-6 \mu _1 \mathcal{R}^2-18,\\
    &d_{2} = \mu _1 r^2 \left(-3 \mu _1 r^2+\mu _1 \mathcal{R}^2+15\right)-6 \mu _1 \mathcal{R}^2-9,\\
    &d_{3} = r^2 \left(\mu _1 \left(3 \mu _1 r^4-16 r^2+\mathcal{R}^2\right)+12\right)+6 \mathcal{R}^2,\\
    &d_{4} = r^2 \left(\mu _1 r^2 \left(7 \mu _1 r^2-\mu _1 \mathcal{R}^2-36\right)+9 \mu _1 \mathcal{R}^2+18\right)+9 \mathcal{R}^2,\\
    &d_{5} = \mu _1 r^2 \left(3 \mu _1 r^2-\mu _1 \mathcal{R}^2-15\right)+6 \mu _1 \mathcal{R}^2+9.
\end{align}

\section{Generic Interior Metric}\label{appendix: solution 2 case II}
Here, we give the full expressions for the solution given in Sec. \ref{sec: CD-TIV}. 
\subsection{Relation among the parameters induced by the junction conditions:}
\begin{equation}
    \mathfrak{D}_{3}= \frac{3 \left(12 \mathfrak{D}_1^3 r_b^4+\mathfrak{D}_1^2 r_b^2 \left(49 \mathfrak{D}_2 r_b^4+13\right)+10 \mathfrak{D}_2 r_b^2 \left(3 \mathfrak{D}_2^2 r_b^8+1\right)+\mathfrak{D}_1 \left(\mathfrak{D}_2 r_b^4 \left(61 \mathfrak{D}_2 r_b^4+30\right)+5\right)\right)}{2 \mathfrak{D}_2 r_b^4 \left(\mathfrak{D}_2 r_b^4 \left(11 \mathfrak{D}_2 r_b^4+36\right)-11\right)+\mathfrak{D}_1 r_b^2 \left(\mathfrak{D}_2 r_b^4 \left(37 \mathfrak{D}_2 r_b^4+14\right)-3\right)+3 \mathfrak{D}_1^2 \left(\mathfrak{D}_2 r_b^8+r_b^4\right)},
\end{equation}
\begin{align}
\mathfrak{D}_{4}= \frac{1}{\mathfrak{d}_{1}}&\left(36 \mathfrak{D}_1^4 r_b^6+3 \mathfrak{D}_1^3 r_b^4 \left(69 \mathfrak{D}_2 r_b^4+25\right)+4 \mathfrak{D}_1^2 r_b^2 \left(\mathfrak{D}_2 r_b^4 \left(71 \mathfrak{D}_2 r_b^4+80\right)+18\right)\right.\nonumber\\ 
&\left.+ 10 \mathfrak{D}_2 r_b^2 \left(\mathfrak{D}_2 r_b^4 \left(23-\mathfrak{D}_2 r_b^4 \left(3 \mathfrak{D}_2 r_b^4+31\right)\right)+3\right)\right.\nonumber\\ &\left.+\mathfrak{D}_1\left(\mathfrak{D}_2 r_b^4 \left(\mathfrak{D}_2 r_b^4 \left(41 \mathfrak{D}_2 r_b^4+129\right)+287\right)+15\right)\right),\\
\mathfrak{D}_{5}= \frac{1}{\mathfrak{d}_{1}}&\left(16 \mathfrak{D}_2^2 r_b^4 \left(\mathfrak{D}_2 r_b^4 \left(2 \mathfrak{D}_2 r_b^4+7\right)-7\right)+8 \mathfrak{D}_1 \mathfrak{D}_2 r_b^2 \left(\mathfrak{D}_2 r_b^4+1\right) \left(5 \mathfrak{D}_2 r_b^4-12\right)\right.\nonumber\\
    &\left.-6 \mathfrak{D}_1^3 \left(5 \mathfrak{D}_2 r_b^6+r_b^2\right) -4 \mathfrak{D}_1^2 \left(\mathfrak{D}_2 r_b^4 \left(10 \mathfrak{D}_2 r_b^4+21\right)+3\right)\right),
\end{align}
where 
\begin{align}
    \mathfrak{d}_{1}= &r_b^2 \left(5 \mathfrak{D}_2 r_b^4+3 \mathfrak{D}_1 r_b^2+1\right) \left(3 \mathfrak{D}_1^2 \left(\mathfrak{D}_2 r_b^6+r_b^2\right)+\mathfrak{D}_1 \left(\mathfrak{D}_2 r_b^4 \left(37 \mathfrak{D}_2 r_b^4+14\right)-3\right)\right.\nonumber\\ 
    &\left.+ 2 \mathfrak{D}_2 r_b^2 \left(\mathfrak{D}_2 r_b^4 \left(11 \mathfrak{D}_2 r_b^4+36\right)-11\right)\right).
\end{align}

\subsection{Ricci scalar:}
\begin{equation}
    R(r)=\frac{2\left(-11 \mathfrak{D}_2^2 \mathfrak{D}_3 \mathfrak{D}_5 r^{12} + \mathfrak{D}_2 \mathfrak{b}_{1} r^{10} - \mathfrak{b}_{2} r^8 - \mathfrak{b}_{3} r^6 + \mathfrak{b}_{4} r^4 - \mathfrak{b}_{5} r^2 - 3 \left(\mathfrak{D}_1-\mathfrak{D}_3+\mathfrak{D}_4\right)\right)}{\left(\mathfrak{D}_2 r^4+\mathfrak{D}_1 r^2+1\right){}^2 \left(\mathfrak{D}_3 r^2+1\right){}^2},
\end{equation}
where
\begin{align}
    \mathfrak{b}_{1} &= \mathfrak{D}_2 \mathfrak{D}_3 \left(\mathfrak{D}_3-7 \mathfrak{D}_4\right)-3 \left(5 \mathfrak{D}_2+6 \mathfrak{D}_1 \mathfrak{D}_3\right) \mathfrak{D}_5,\\
    \mathfrak{b}_{2} &= \mathfrak{D}_2 \mathfrak{D}_3 \left(\mathfrak{D}_2-2 \mathfrak{D}_1 \mathfrak{D}_3\right) + 11 \mathfrak{D}_2 \left(\mathfrak{D}_2+\mathfrak{D}_1 \mathfrak{D}_3\right) \mathfrak{D}_4+25 \mathfrak{D}_1 \mathfrak{D}_2 \mathfrak{D}_5 \nonumber\\
    &+6 \left(\mathfrak{D}_1^2+3 \mathfrak{D}_2\right) \mathfrak{D}_3 \mathfrak{D}_5,\\
    \mathfrak{b}_{3} &= 6 \mathfrak{D}_2^2+2 \left(-\mathfrak{D}_3^2+6 \mathfrak{D}_4 \mathfrak{D}_3+9 \mathfrak{D}_1 \mathfrak{D}_4+12 \mathfrak{D}_5\right) \mathfrak{D}_2\nonumber\\
    &+\mathfrak{D}_1 \left(10 \mathfrak{D}_3 \mathfrak{D}_5+\mathfrak{D}_1 \left(-\mathfrak{D}_3^2+3 \mathfrak{D}_4 \mathfrak{D}_3+9 \mathfrak{D}_5\right)\right),\\
    \mathfrak{b}_{4} &= 2 \left(\mathfrak{D}_3-3 \mathfrak{D}_4\right) \mathfrak{D}_1^2-\left(-2 \mathfrak{D}_3^2+5 \mathfrak{D}_4 \mathfrak{D}_3+9 \mathfrak{D}_2+15 \mathfrak{D}_5\right) \mathfrak{D}_1-2 \mathfrak{D}_2 \left(\mathfrak{D}_3+9 \mathfrak{D}_4\right) \nonumber\\
    &-3 \mathfrak{D}_3 \mathfrak{D}_5,\\
    \mathfrak{b}_{5} &= (2 \mathfrak{D}_1^2+\left(10 \mathfrak{D}_4-4 \mathfrak{D}_3\right) \mathfrak{D}_2+\mathfrak{D}_3 \left(\mathfrak{D}_4-\mathfrak{D}_3\right)+5 \mathfrak{D}_5.
\end{align}

\subsection{Total energy density:}
\begin{equation}
    \mu^{\text{tot}}= \frac{-3 r^4 \mathfrak{D}_{3} \mathfrak{D}_{5}+r^2 \left(\mathfrak{D}_{3}{}^2-\mathfrak{D}_{3} \mathfrak{D}_{4}-5 \mathfrak{D}_{5}\right)+3 \mathfrak{D}_{3}-3 \mathfrak{D}_{4}}{\left(r^2 \mathfrak{D}_{3}+1\right){}^2}
\end{equation}

\subsection{Total radial pressure:}
\begin{align}
    p_{r}^{\text{tot}} = &\frac{1}{\left(r^2 \mathfrak{D}_{3}+1\right) \left(r^4 \mathfrak{D}_{2}+r^2 \mathfrak{D}_{1}+1\right)}\left(5 r^6 \mathfrak{D}_{2} \mathfrak{D}_{5}+r^4 (3 \mathfrak{D}_{1} \mathfrak{D}_{5}-\mathfrak{D}_{2} \mathfrak{D}_{3}+5 \mathfrak{D}_{2} \mathfrak{D}_{4})\right.\nonumber\\
    &\left.+r^2 (-\mathfrak{D}_{1} \mathfrak{D}_{3}+3 \mathfrak{D}_{1} \mathfrak{D}_{4}+4 \mathfrak{D}_{2}+\mathfrak{D}_{5})+2 \mathfrak{D}_{1}-\mathfrak{D}_{3}+\mathfrak{D}_{4}\right)
\end{align}

\subsection{Total orthogonal pressure}
\begin{align}
    p_{\perp}^{\text{tot}} = &\frac{1}{\left(r^2 \mathfrak{D}_{3}+1\right){}^2 \left(r^4 \mathfrak{D}_{2}+r^2 \mathfrak{D}_{1}+1\right){}^2}\left(7 r^{12} \mathfrak{D}_{2}{}^2 \mathfrak{D}_{3} \mathfrak{D}_{5} + r^{10} \mathfrak{g}_{1} + r^8 \mathfrak{g}_{2} + r^6 \mathfrak{g}_{3} + r^4 \mathfrak{g}_{4} \right.\nonumber\\
    &\left.+ r^2 \mathfrak{g}_{5} + 2 \mathfrak{D}_{1}-\mathfrak{D}_{3}+\mathfrak{D}_{4}\right),
\end{align}
where
\begin{align}
    \mathfrak{g}_{1} &= \mathfrak{D}_{2} (11 \mathfrak{D}_{1} \mathfrak{D}_{3} \mathfrak{D}_{5}+4 \mathfrak{D}_{2} \mathfrak{D}_{3} \mathfrak{D}_{4}+10 \mathfrak{D}_{2} \mathfrak{D}_{5}),\\
    \mathfrak{g}_{2} &= 3 \mathfrak{D}_{3} \mathfrak{D}_{5} \left(\mathfrak{D}_{1}{}^2+4 \mathfrak{D}_{2}\right)+6 \mathfrak{D}_{1} \mathfrak{D}_{2} \mathfrak{D}_{3} \mathfrak{D}_{4}+16 \mathfrak{D}_{1} \mathfrak{D}_{2} \mathfrak{D}_{5}+\mathfrak{D}_{2}{}^2 (\mathfrak{D}_{3}+7 \mathfrak{D}_{4}),\\
    \mathfrak{g}_{3} &= \mathfrak{D}_{2} (\mathfrak{D}_{1} \mathfrak{D}_{3}+11 \mathfrak{D}_{1} \mathfrak{D}_{4}+8 \mathfrak{D}_{3} \mathfrak{D}_{4}+16 \mathfrak{D}_{5})+\mathfrak{D}_{1} (\mathfrak{D}_{1} \mathfrak{D}_{3} \mathfrak{D}_{4}+5 \mathfrak{D}_{5} (\mathfrak{D}_{1}+\mathfrak{D}_{3}))+4 \mathfrak{D}_{2}{}^2,\\
    \mathfrak{g}_{4} &= -\mathfrak{D}_{1}{}^2 (\mathfrak{D}_{3}-3 \mathfrak{D}_{4})+2 \mathfrak{D}_{1} (3 \mathfrak{D}_{2}+\mathfrak{D}_{3} \mathfrak{D}_{4}+4 \mathfrak{D}_{5})+4 \mathfrak{D}_{2} (\mathfrak{D}_{3}+3 \mathfrak{D}_{4})+\mathfrak{D}_{3} \mathfrak{D}_{5},\\
    \mathfrak{g}_{5} &= \mathfrak{D}_{1}{}^2-\mathfrak{D}_{1} \mathfrak{D}_{3}+5 \mathfrak{D}_{1} \mathfrak{D}_{4}+8 \mathfrak{D}_{2}+2 \mathfrak{D}_{5}.
\end{align}

\subsection{Total isotropic pressure}
\begin{equation}
    p^{\text{tot}} = \frac{19 r^{12} \mathfrak{D}_{2}{}^2 \mathfrak{D}_{3} \mathfrak{D}_{5} + r^{10} \mathfrak{h}_{1} + r^8 \mathfrak{h}_{2} + r^6 \mathfrak{h}_{3} + r^4 \mathfrak{h}_{4} + r^2 \mathfrak{h}_{5} + 6 \mathfrak{D}_{1}-3 \mathfrak{D}_{3}+3 \mathfrak{D}_{4}}{3 \left(r^2 \mathfrak{D}_{3}+1\right){}^2 \left(r^4 \mathfrak{D}_{2}+r^2 \mathfrak{D}_{1}+1\right){}^2}
\end{equation}
where
\begin{align}
    \mathfrak{h}_{1} &= \mathfrak{D}_{2} (5 \mathfrak{D}_{5} (6 \mathfrak{D}_{1} \mathfrak{D}_{3}+5 \mathfrak{D}_{2})-\mathfrak{D}_{2} \mathfrak{D}_{3} (\mathfrak{D}_{3}-13 \mathfrak{D}_{4})),\\
    \mathfrak{h}_{2} &= \mathfrak{D}_{5} \left(9 \mathfrak{D}_{1}{}^2 \mathfrak{D}_{3}+40 \mathfrak{D}_{1} \mathfrak{D}_{2}+30 \mathfrak{D}_{2} \mathfrak{D}_{3}\right)+\mathfrak{D}_{2} (\mathfrak{D}_{2} (5 \mathfrak{D}_{3}+19 \mathfrak{D}_{4})-2 \mathfrak{D}_{1} \mathfrak{D}_{3} (\mathfrak{D}_{3}-10 \mathfrak{D}_{4})),\\
    \mathfrak{h}_{3} &= \mathfrak{D}_{2} \left(6 \mathfrak{D}_{1} (\mathfrak{D}_{3}+5 \mathfrak{D}_{4})-2 \mathfrak{D}_{3}{}^2+22 \mathfrak{D}_{3} \mathfrak{D}_{4}+38 \mathfrak{D}_{5}\right)\nonumber\\ 
    &+\mathfrak{D}_{1} \left(\mathfrak{D}_{1} \left(-\mathfrak{D}_{3}{}^2+5 \mathfrak{D}_{3} \mathfrak{D}_{4}+13 \mathfrak{D}_{5}\right)+14 \mathfrak{D}_{3} \mathfrak{D}_{5}\right)+12 \mathfrak{D}_{2}{}^2,\\
    \mathfrak{h}_{4} &= -\mathfrak{D}_{1}{}^2 (\mathfrak{D}_{3}-9 \mathfrak{D}_{4})+2 \mathfrak{D}_{1} \left(9 \mathfrak{D}_{2}-\mathfrak{D}_{3}{}^2+4 \mathfrak{D}_{3} \mathfrak{D}_{4}+10 \mathfrak{D}_{5}\right)+10 \mathfrak{D}_{2} (\mathfrak{D}_{3}+3 \mathfrak{D}_{4})+3 \mathfrak{D}_{3} \mathfrak{D}_{5},\\
    \mathfrak{h}_{5} &= 4 \mathfrak{D}_{1}{}^2-2 \mathfrak{D}_{1} (\mathfrak{D}_{3}-7 \mathfrak{D}_{4})+20 \mathfrak{D}_{2}+\mathfrak{D}_{3} (\mathfrak{D}_{4}-\mathfrak{D}_{3})+5 \mathfrak{D}_{5}.
\end{align}

\end{document}